\documentclass[a4paper, 11pt]{article}
\makeatletter
\let\@fnsymbol\@arabic
\makeatother
\usepackage{graphicx} 
\usepackage[export]{adjustbox}
\usepackage[english]{babel}\usepackage{booktabs}	
\usepackage{bbm}
\usepackage{tikz}
\usepackage[T1]{fontenc}
\usepackage{lmodern}
\usepackage{chngcntr}
\usepackage{stackrel}
\usepackage{enumerate}
\usepackage{float} 		
\usepackage[round]{natbib}
\usepackage{epstopdf}
\usepackage[toc,page]{appendix} 
\usepackage[intlimits]{amsmath}
\usepackage{amsfonts}
\usepackage{rotating} 	
\usepackage{graphicx} 	
\usepackage{ae,aecompl}
\usepackage{appendix}
\usepackage{amsmath,amsthm,amssymb,enumerate,mathrsfs}
\usepackage{epsfig}

\usepackage[hidelinks]{hyperref}
\hypersetup{
    colorlinks=true,
    linkcolor=blue,
    citecolor=blue,
    urlcolor=blue
}

\usepackage{caption}
\usepackage{multirow}
\usepackage{epsfig}    
\usepackage{lscape}    
\usepackage{caption}
\usepackage{subcaption}
\usepackage{setspace}
\usepackage{url}
\usepackage[a4paper]{geometry}
\usepackage[flushleft]{threeparttable}

\normalsize
\newtheorem{Proposition}{Proposition}

\newtheorem{Property}[Proposition]{Property}

\newtheorem{hypothesis}{Hypothesis}

\newtheorem{recommendation}{Recommendation}

\numberwithin{equation}{section}
\numberwithin{Proposition}{section}
\numberwithin{table}{section}
\usepackage{setspace}
\usepackage{substitutefont}
\substitutefont{TS1}{aer}{lmr}
  
\title{Backtesting Expected Shortfall: Accounting for both duration and severity with bivariate orthogonal polynomials}
\author{Sullivan Hu\'e\thanks{Aix-Marseille University (Aix-Marseille School of Economics), CNRS $\&$ EHESS. Email: sullivan.hue@univ-amu.fr} \and Christophe Hurlin\footnote{University of Orléans (LEO) and Institut Universitaire de France. E-mail: christophe.hurlin@univ-orleans.fr} \and Yang Lu\thanks{Concordia University, Mont\'real, Canada. Email: yang.lu@concordia.ca}}

\begin{document}
\maketitle
\begin{abstract}
\noindent We propose an original two-part, duration-severity approach for backtesting Expected Shortfall (ES). While Probability Integral Transform (PIT) based ES backtests have gained popularity, they have yet to allow for separate testing of the frequency and severity of Value-at-Risk (VaR) violations. This is a crucial aspect, as ES measures the average loss in the event of such violations. To overcome this limitation, we introduce a backtesting framework that relies on the sequence of inter-violation durations and the sequence of severities in case of violations. By leveraging the theory of (bivariate) orthogonal polynomials, we derive orthogonal moment conditions satisfied by these two sequences. Our approach includes a straightforward, model-free Wald test, which encompasses various unconditional and conditional coverage backtests for both VaR and ES. This test aids in identifying any mis-specified components of the internal model used by banks to forecast ES. Moreover, it can be extended to analyze other systemic risk measures such as Marginal Expected Shortfall. Simulation experiments indicate that our test exhibits good finite sample properties for realistic sample sizes. Through application to two stock indices, we demonstrate how our methodology provides insights into the reasons for rejections in testing ES validity.
\end{abstract}

\vspace{1em}

\noindent \baselineskip=0.9\normalbaselineskip \textbf{Acknowledgements}: We thank the NSERC (through grants RGPIN-2021-04144 and DGECR-2021-00330), the Institut Universitaire de France (IUF), the French National Research Agency (MLEforRisk ANR-21-CE26-0007), and the Excellence Initiative of Aix-Marseille University (A*MIDEX), for supporting our research. We thank seminar participants at various universities for helpful comments. 

\newpage

 \section{Introduction}

The introduction of the Basel III Accord marked a significant shift in market risk measurement for banks, replacing Value-at-Risk (VaR) with Expected Shortfall (ES).\footnote{The Basel III accord was published by the Basel Committee on Banking Supervision in 2010. The transition of the quantitative risk metrics system from VaR to ES for market risks was accompanied by a lowering of the confidence level from 99\% to 97.5\%.} Under the Accord, banks are mandated to report their daily ES for all market risk exposures, determining their level of regulatory capital holdings. Consequently, backtesting ES has become essential for banks.

\medskip

Three main approaches to ES backtesting can be distinguished \citep{deng2021backtesting}. The first one is the multi-level VaR tests \citep{colletaz2013risk, emmer2015best, wied2016evaluating, kratz2018multinomial, couperier2019backtesting}. This literature approximates ES by a Riemann sum of VaRs at several levels deeper in the tail, and hence transforms the ES backtesting problem into backtesting multiple VaRs. The downside of this approach is that it requires banks to report their VaRs at multiple levels, a practice not currently implemented in the industry. Furthermore, even if such reporting becomes mandatory, backtesting VaRs at extreme levels is highly challenging due to the scarcity of VaR violations (i.e., exceedances). A second approach explores the joint elicitability of the couple (ES,VaR) to test these two risk measures \textit{simultaneously}. This family includes  regression type tests based on the joint score \citep{nolde2017elicitability, dimitriadis2019joint, patton2019dynamic, bayerregression, dimitriadis2023encompassing}, tests of moment conditions derived from the identification function
\citep{nolde2017elicitability, barendsebacktesting}, as well as the e-value based test \citep{wang2022backtesting}. Even though this approach is promising, we do not follow this literature for several reasons. Firstly, many tests, such as score tests and e-value tests, involve a single joint test. In case of rejection of the null hypothesis, risk managers are unable to discern whether VaR or ES is responsible for the outcome. Secondly, numerous tests do not allow for separate assessments of the unconditional coverage (UC) and independence (IND) assumptions. Such separate tests have become common practice in the backtesting literature, offering banks insights on which part of the internal model needs improvement. Thirdly, regression-based tests assume a linear model under the alternative hypothesis, which may be overly restrictive. Lastly, identification function-based tests typically assess only a small number of conditional moment conditions due to the curse of dimensionality.\footnote{See also \cite{hoga2023monitoring} for further discussions on the pros and cons of this approach compared to the PIT approach described in the next paragraph.} 

\medskip

Finally, a growing literature leverages the Probability Integral Transform (PIT), also known as standardization. The PIT provides, at each date, the realized quantile, \textit{i.e.}, the level of the predictive quantile matching the realized return.  This framework directly extends the seminal VaR backtesting methodology proposed by \cite{christoffersen1998evaluating}. Indeed, comparing the PIT of the return series with a fixed level, such as 95\%, generates a binary violation process. Since \cite{christoffersen1998evaluating}, this process has been the cornerstone of VaR backtests. The approach based on PIT is a major improvement compared to the violation-only approach, since the PIT reveals not only the frequency of VaR violations, but also the tail distribution beyond the VaR. Moreover, PIT reporting has recently also become the industry standard \citep{gordy2020spectral} thanks to its ability of ensuring a high level of risk reporting by banks, without requesting full disclose of their internal models.  However, this research area is still nascent, with most existing PIT-based backtests addressing UC only  \citep{acerbi2002spectral,kerkhof2004backtesting,costanzino2015backtesting,loser2018new}. To our knowledge, \cite{du2017backtesting} is the first to propose conditional coverage (CC) tests of ES through PIT, and this approach has recently also been adopted by \cite{hoga2023monitoring} and \cite{Du2023}. Our paper also follows this PIT approach.

\medskip

A key ingredient of the aforementioned PIT based tests is the  cumulative violation process:
\begin{equation}
    \label{pit}
    H_t(\alpha)=\frac{1}{\alpha}(\alpha-u_t)\mathbbm{1}(u_t \leq \alpha),
\end{equation}
where $\alpha \in ]0,1[$ denotes the coverage level and $u_t$ is the conditional PIT given past information. Under the null hypothesis that the bank has an accurate model for its Profit and Loss (P\&L), $u_t$ is i.i.d. and uniformly distributed over $[0,1]$. If the PIT exceeds $\alpha$, then $H_t(\alpha)$ is equal to zero, indicating no VaR violation. If instead $u_t$ is smaller than $\alpha$, then $H_t(\alpha)$ is positive, measuring the ``severity" of the VaR violation. Under the null, the conditional distribution of $H_t(\alpha)$ given $u_t<\alpha$ is also uniform. In their seminal article, \cite{du2017backtesting} propose testing $\mathbb{E}[H_t(\alpha)]=\alpha/2$ for the UC test and the lack of serial correlation of the process $(H_t(\alpha))$ for the IND test. However, these tests do not differentiate between the two components of the support of $H_t(\alpha)$, that are the point mass at zero, and the continuous component on $]0,1]$ which addresses the specification of the tail risk beyond the VaR. Thus, solely considering the expectation (resp. autocorrelation) of $H_t(\alpha)$ fails to adequately characterize its marginal distribution (resp. serial dependence). Additionally, since $H_t(\alpha)$ cannot be written as a simple function of the daily ES, a small number of moment conditions on $H_t(\alpha)$ may not be strong enough to ensure the correct specification of the ES. This means that we can find wrong models that mis-specifies both the discrete and the continuous components, but not the few tested moment conditions. In practice, such flawed models may produce inaccurate ES forecasts, yet pass existing backtests due to their limited power against these alternatives.

\medskip

In this paper, we use a joint orthogonal polynomial expansion to derive moment conditions for these two components. This approach enables us to incorporate as many moment conditions as desired, enhancing the test's ability to detect mis-specifications. To tackle the two-component support of $H_t(\alpha)$, we first transform the process $(H_t(\alpha))$ into two sequences, one representing the VaR violations and the other representing the severities of these violations. The violation process, defined as:
\begin{equation}
    \label{violation}
    h_t(\alpha)=\mathbbm{1}(H_t(\alpha)>0)=\mathbbm{1}(u_t<\alpha),
\end{equation}
is observable at each date $t$. The sequence of severities, observable only when a VaR violation occurs, is equal to $H_t(\alpha)$ at these  dates. This separation of the two components enables more efficient use of information. Specifically, we show how incorporating the severity variable improves the power of our CC VaR test. Similarly, by incorporating the violation sequence into the CC test of ES model, the power of the CC ES test can be improved as well. In other words, this ``two-part" approach not only allows for independent validation of the specification of VaR and ES, but also increases the power of the CC test for both models. Indeed, a common criticism of many VaR backtests is that by focusing solely on the binary violation process, much information is overlooked \citep{gordy2020spectral}.

\medskip

Our second contribution is the introduction of inter-violation durations into the literature on ES. The sequence of durations has a one-to-one relationship with the violation process. Similarly, the usual i.i.d. Bernoulli$(\alpha)$ hypothesis of the latter is equivalent to the assumption that durations are i.i.d., following geometric distribution with probability parameter $\alpha$. In the VaR literature, duration-based tests have proven to be a significant competitor to traditional violation-based tests \citep{christoffersen2004backtesting, Haas2006,berkowitz2011evaluating, candelon2011backtesting, ziggel2014new, pelletier2016geometric} when it comes to the power of the tests. In ES backtesting, the use of durations becomes particularly relevant due to the analysis of dependence between violation and severity sequences. While the violation process is observed daily, severities are only observed when a violation occurs. Transforming the violation sequence into durations resolves this issue, as there are as many durations as there are severities, facilitating the characterization of inter-dependence between violations and severities. Despite the aforementioned attempts, duration-based testing is still in its early stages and has yet to be applied to ES backtesting. As noted by \cite{candelon2011backtesting}, a challenge of the duration-based tests (for VaR) is to develop formal separate tests for $(i)$ the UC, $(ii)$ the IND assumption, and eventually $(iii)$ the CC assumption within a \textit{unified} framework. In this paper, we introduce the first duration-based ES backtest and address these concerns by utilizing the theory of (bivariate) orthogonal polynomials. 

\medskip

Third, we are the first to propose an application of bivariate orthogonal polynomials in Finance. Previously, the use of orthogonal polynomials in financial application was primarily restricted to univariate distributions, particularly focusing on the normal distribution and its associated Hermite polynomials \citep{corrado1996skewness, ait2002maximum, bontemps2005testing, Bontemps_Meddahi_2012}. In our study, we consider bivariate orthogonal polynomials, which allow us to test independence between two variables with known marginal distributions, such as the serial and mutual independence between duration and severity variables. We employ Legendre orthogonal polynomials, associated to the uniform distribution (i.e., the distribution of the PIT under the null), to test the specification of the severity variable. Additionally, following the approach of \citet{candelon2011backtesting}, we utilize Meixner orthogonal polynomials, associated to the geometric distribution, to test the marginal distributions of the durations between two consecutive VaR violations. Our orthogonal polynomial-based tests offer several advantages. Firstly, they are model-free and non-parametric, as they do not impose parametric assumptions (e.g., Markov, Weibull, EACD) under the null or alternative hypothesis. This is particularly advantageous, as it can be challenging to find sufficiently general parametric alternative hypotheses.\footnote{For instance, in \citet{christoffersen2004backtesting}, the alternative hypothesis is that the durations follow the Weibull distribution. But the Weibull distribution is continuous and is thus not appropriate for discrete durations.} Secondly, the person in charge of model validation has the freedom to choose the number or type (e.g., serial, cross-sectional, marginal) of moment conditions to test. On one hand, testing more moments increases the likelihood of detecting violations by a wrong model, thereby ensuring that our test has power against a wide range of alternatives. On the other hand, by selecting appropriate types of moment conditions, we can recover various unconditional coverage (UC) and conditional coverage (CC) tests for both VaR and ES as special cases. These two properties suggest a straightforward method for risk managers to identify potential mis-specified components of the internal model. They can begin with a test involving many conditions, and if it results in rejection, they can then consider nested tests with fewer, nested moment conditions. Thirdly, orthogonal polynomials ensure that the covariance matrix of the sample moments is the identity matrix, eliminating the need for estimation. As a result, our (Wald) test does not suffer from the usual curse of dimensionality that might arise from estimating the asymptotic covariance matrix, even when a large number of moments are involved. Indeed, \citet{guo2001testing}, \citet{nolde2017elicitability} and \citet{barendsebacktesting} report that when such covariance matrices have to be estimated, the Wald test can suffer from significant size distortion, as well as low power for reasonable sample size.

\medskip

Another advantage of our backtesting procedure is its applicability to several prominent systemic risk measures, such as Marginal Expected Shortfall (MES) or Systemic Risk measure (SRISK) introduced by \citet{acharya2017measuring} and \citet{Brownlees_Engle_2017}, respectively. The definition of these systemic risk measures also involves conditioning, akin to VaR violations in the global market, similar to the definition of ES. Thus, duration and severity variables can be readily constructed, enabling similar orthogonal polynomial-based backtests.  

\medskip

The paper is organized as follows. Section 2 introduces the methodology. Section 3 reports results of Monte Carlo experiments. Section 4 extends the framework to backtest systemic risk measures such as Marginal Expected Shortfall. Section 5 applies the methodology to stock index data. Section 6 concludes. Technical details are gathered in Appendices.

\section{The methodology}

\subsection{The state-of-art framework}

Let $y_t$ denote the P\&L (or return) of the bank at time $t$, $\Omega_{t-1}$ the information set available at time $t-1$, and $G_t(.,\Omega_{t-1})$ the conditional distribution of $y_t$ given $\Omega_{t-1}$, which is assumed to be continuous. The conditional VaR at a coverage level $\alpha\in]0,1[$ is defined as the negative of the $\alpha^{th}$-quantile of the P\&L's conditional distribution: 
\begin{equation}
    \label{defvar}
    VaR_t(\alpha)=-G_t^{-1}(\alpha,\Omega_{t-1}),  \hspace{1cm}       \mathbb{P}[y_t\leq-VaR_t(\alpha)|\Omega_{t-1}]=\alpha.
\end{equation}
It is often convenient to introduce $u_t=G_t(y_t,\Omega_{t-1})$, i.e., the (conditional) PIT of $y_t$. The PIT $u_t$ provides the level of the predictive quantile that matches the realized return $y_t$ at time $t$. Then, we define the VaR violation process as follows:
\begin{equation}
    \label{violation_VaR}
    h_t(\alpha)=\mathbbm{1}(y_t \leq -VaR(\alpha))=\mathbbm{1}(u_t \leq \alpha). 
\end{equation}
That is, given a level $\alpha$, a VaR violation occurs, when the realized quantile of the return is below $\alpha$. Thus the second equation in Eq. \eqref{defvar} becomes:
\begin{equation}
    \label{iid}
    \mathbb{P}[h_t(\alpha)=1|\Omega_{t-1}]=\alpha.
\end{equation}
When the conditional distribution $G_t(.,\Omega_{t-1})$ is well specified, indicating the validity of the bank's internal model, the violation process $(h_t(\alpha))$ is i.i.d., and the VaR forecast is said to satisfy a conditional coverage (CC) assumption. This property forms the foundation of most VaR backtests since \cite{Kupiec1995} and \cite{christoffersen1998evaluating}. The CC assumption implies an unconditional coverage (UC) condition on the marginal probability, such that:
\begin{equation}
\label{ucvar}
    \mathbb{P}[h_t(\alpha)]=1,
\end{equation}
The CC assumption also means that VaR violations are independent (IND), implying the lack of correlation condition:
  \begin{equation}
      \label{indvar}
      \text{Corr}[h_t(\alpha), h_{t-k}(\alpha)]=0, \qquad \forall k \in  \mathbb{Z}.
  \end{equation} 

\medskip

The VaR has two main limitations. Firstly, it is not a coherent risk measure \citep{Artzner1999}, as it does not satisfy the subadditivity axiom, according to which the risk measure should not penalize diversification. Secondly, VaR does not account for the severity of the loss, in case it exceeds the VaR threshold. This is why, since Basel III, the Basel Committee on Banking Supervision (BCBS) has replaced VaR by ES for measuring market risk. The ES measures the expected loss incurred on a portfolio given that the loss exceeds VaR. Formally, the $\alpha$-level ES is defined as the negative of the conditional expected return given that the return is less than $-VaR_t(\alpha)$:
\begin{align}
    \label{conditionallosses}
    ES_t(\alpha)&=-\mathbb{E}[y_t|\Omega_{t-1},y_t<-VaR_t(\alpha)]\\
    &=\frac{1}{\alpha} \int_0^{\alpha} VaR_t(u) \,du \label{equation_ES}.
\end{align}
Various models are used by banks to forecasts VaR and ES, see e.g., \citet{McN05,Perignon2010,patton2019dynamic}. 

\medskip

When it comes to backtesting ES, Eq. \eqref{equation_ES} can been used to approximate the ES by a Riemann summation \citep{colletaz2013risk, kratz2018multinomial, couperier2019backtesting}. However, this approach has several downsides. Firstly, the estimation of $VaR_t(u)$ for extremely small $u$ could be very challenging. Secondly, and most importantly, in practice, the validation team in charge of the backtesting may not have access to all $VaR_t(u)$, for all the coverage levels $u \in ]0,\alpha]$. Instead, oftentimes, only $VaR_t(\alpha)$ for one given value $\alpha$ (say, 5 \%) is available. On the other hand, the PIT series $u_t$ is often observable, as reporting PIT has recently become standard in the financial industry \citep{gordy2020spectral},  thanks to its advantage of shielding banks from disclosing their internal models. This motivates the introduction by \cite{costanzino2015backtesting}, \cite{du2017backtesting}, \cite{loser2018new}, \cite{du2017backtesting}, \cite{hoga2023monitoring}, \cite{Du2023} of PIT-based ES backtests that rely on the \textit{cumulative} violation process $H_t(\alpha)$ defined as:
\begin{equation}
    H_t(\alpha)=\frac{1}{\alpha}(\alpha-u_t)\mathbbm{1}(u_t \leq \alpha),
\end{equation}
\cite{du2017backtesting} show that $H_t(\alpha)$ can be equivalently written as:
\begin{equation}  
    \label{definitionviolation}
    H_t(\alpha)=\frac{1}{\alpha} \int_{0}^{\alpha} h_t(u) \,du.\end{equation}
By Fubini's Theorem, the martingale difference sequence (mds) property of the process $h_t(\alpha)-\alpha$ is preserved by integration, which means that $H_t(\alpha)-\alpha/2$ is also a mds. By exploiting this property, \cite{du2017backtesting} propose an intuitive framework, similar to that used for the VaR-backtests, to test the UC condition on the ES given by:
\begin{equation}
    \label{ucdu}
    \mathbb{E}[H_t(\alpha)]=\frac{\alpha}{2},
\end{equation}
They also propose a test for the IND assumption for the ES based on the lack of correlation of the sequence $(H_t(\alpha))$:
\begin{equation}
    \label{inddu}
    \text{Corr}[H_t(\alpha),H_{t-k}(\alpha)]=0, \qquad \forall k\in\mathbb{Z},
\end{equation}
This approach has recently been adopted by \cite{hoga2023monitoring}, while \cite{Du2023} propose an improvement of condition \eqref{inddu}, which is tailored to ES forecasts obtained by Historical Simulation (HS) and Filtered Historical Simulation (FHS) methods. This approach has also been applied to backtest systemic risk measures such as MES or SRISK by \cite{banulescu2021backtesting}.

\subsection{The limitations and a two-part frequency-severity approach}\label{limitations}

The above approach is based on two elegant analogies: 
\begin{enumerate}[$i)$]

    \item The UC \eqref{ucdu} and IND \eqref{inddu} conditions for $H_t(\alpha)$ in ES backtesting are analogous to the UC \eqref{ucvar} and IND \eqref{indvar} conditions for the violation process $h_t(\alpha)$ used in VaR backtesting.
    
    \item The cumulative violation $H_t(\alpha)$ defined in \eqref{definitionviolation} can be considered as a violation "counterpart" of the ES definition \eqref{equation_ES}.

\end{enumerate}

These two analogies, however, have to be used with care for two reasons. First, regarding analogy $i)$, since the ES is the average loss in case of VaR violation, ES backtests should ideally check that both the frequency of the violations, and the average loss upon violation, are well specified. Similarly, the support of $H_t(\alpha)$ has two components: one discrete and the other continuous, which require separate treatment. This is unlike VaR backtesting, where the violation sequence $h_t(\alpha)$ is binary, and Eqs. \eqref{ucvar} and \eqref{indvar} fully characterize the marginal and pairwise joint distribution of the process $(h_t(\alpha))$. As a consequence, conditions \eqref{ucdu} and \eqref{inddu} are non sufficient for $(u_t)$ to be i.i.d. and uniform. For instance, simple algebra shows that condition \eqref{ucdu} can be written as:
\begin{equation}
    \label{equivalent}
    \mathbb{P}[u_t< \alpha] \times \mathbb{E}\Big[\frac{\alpha-u_t}{\alpha} \Bigm| u_t< \alpha\Big]=\frac{\alpha}{2},
\end{equation}
or equivalently:
\begin{equation}
    \label{gcondition}
    \int_0^\alpha g(u)(\alpha-u) \mathrm{d}u=\frac{\alpha^2}{2},
\end{equation}
or
\begin{equation}
  \label{gcondition2}
      \int_0^\alpha [g(u)-1](\alpha-u) \mathrm{d}u=0,
\end{equation}
where $g$ is the marginal density of the PIT sequence $(u_t)$. Let us consider the Hilbert space $\mathcal{H}$ of square integrable functions on $(0, \alpha)$, on which $<f, g>:=\int_0^\alpha f(u)g(u)\mathrm{d}u$ defines an inner product. Then condition \eqref{gcondition2} is equivalent to function $g-1$ being orthogonal to function $\alpha-u$ in space $\mathcal{H}$. In other words, $g-1$ belongs to an infinite dimensional hyperplane of $\mathcal{H}$. For instance, one simple example of $g$ is:
\begin{equation}
\label{candidateg}
g(u):=1+b [1+\frac{3}{2\alpha} (\alpha-u)],
\end{equation}
where $b$ is any real number such that $g$ defined in \eqref{candidateg} remains nonnegative on $(0, \alpha)$. Here, the factor $\frac{3}{2\alpha}$ is suitably chosen such that condition \eqref{gcondition2} is satisfied.  
In particular, from this special example, we can see that orthogonality condition \eqref{gcondition2} does not require $\mathbb{P}[u_t< \alpha]= \alpha$, which is the UC condition for VaR, nor does it require $\mathbb{E}\Big[\frac{\alpha-u_t}{\alpha} \bigm| u_t< \alpha\Big]= 1/2$, which concerns the tail risk beyond VaR. 
This example can be further extended to a dynamic model satisfying the IND condition as well. Let us for instance assume that the daily return follows a GARCH-type location-scale model with $y_t= m_t+\sigma_t \epsilon_t$ where $m_t$ and $\sigma_t$ are the conditional mean and volatility, and the $\text{i.i.d.}$ innovation $(\epsilon_t)$ has a cumulative distribution function (CDF) $G(.)$. A hypothetical bank correctly specifies both $m_t$ and $\sigma_t$, but uses a wrong CDF, denoted $F(.)$, for the innovation $\epsilon_t$. Then the sequence of PIT computed by this bank is still i.i.d. (hence this sequence of PIT satisfies \eqref{inddu}). However, its distribution is no longer uniform, but has the CDF $G \circ F^{-1}$. Therefore, for this PIT sequence to also satisfy condition \eqref{ucdu}, it suffices for its corresponding density to be satisfy eq.\eqref{gcondition2}. In other words, a wrong model can satisfy both \eqref{ucdu} and \eqref{inddu} while potentially systematically under-estimating the ES. This suggests that $\mathbb{P}[u_t< \alpha]=\alpha$ and $\mathbb{E}\Big[\frac{\alpha-u_t}{\alpha} \bigm| u_t< \alpha\Big]=1/2$ both need to be tested, to address the UC coverage of frequency and the severity-given-violation of the VaR violations, respectively.\footnote{Note that this limitation of testing a single process (in this case, the cumulative violation) is not unique to the test proposed by  \cite{du2017backtesting}. For example, the popular test of \citet{acerbi2014back} relies on the condition:
\begin{equation}
    \label{acerbi}
    \mathbb{E}\Big[\frac{y_t}{ES_t(\alpha)}+1  \Bigm| y_t< -VaR_t(\alpha)\Big]=0.
\end{equation}
If both $VaR_t(\alpha)$ and $ES_t(\alpha)$ are mis-specified, such that $VaR_t(\alpha)$ computed by this bank corresponds to the true VaR at level $\beta=F \circ G^{-1}(\alpha) \neq \alpha$, whereas $ES_t(\alpha)$ corresponds to the true ES at the same level $\beta$. By the definition of $ES_t(\beta)$, Eq. \eqref{acerbi} is automatically satisfied by this wrong model. Once again, when the test statistic depends on the specification of both the VaR and the ES, but only one condition is tested, the effects of several mis-specifications could potentially cancel out, leading to loss of power against certain alternatives.} The same can be said for the IND coverage assumption. 

\medskip
 
A second limitation concerns analogy $ii)$, and more generally all PIT-based backtests. For VaR backtesting, the violation indicator $h_t(\alpha)=\mathbbm{1}(y_t<-Var_t(\alpha))$ is a simple function of VaR and realized return $y_t$. However, for ES backtesting, the cumulative violation $H_t(\alpha)=\frac{1}{\alpha} \int_{0}^{\alpha} h_t(u) \,du$ cannot be expressed as a simple, linear function of $ES_t(\alpha)$, $VaR_t(\alpha)$, and  $y_t$. Because of this nonlinearity, one condition written on the linear expectation of $H_t(\alpha)$, say $\mathbb{E}[H_t(\alpha)]=\alpha/2$, is no longer a linear expectation type condition on $ES_t$. Similarly, the lack of (linear) autocorrelation of $H_t(\alpha)$ is not sufficient to insure that this sequence is i.i.d. 

\medskip

In summary, there are two important aspects in which existing tests could be improved:
\begin{enumerate}[$i)$]
    \setlength{\itemsep}{0pt}
    \item To introduce at least two processes capturing the frequency and the severity-given-violation of the VaR violations, respectively.
    \item To increase the number of (nonlinear) conditions to test.
\end{enumerate}
To address task $i)$, we note that the sequence $(H_t(\alpha)), t=1,...,T$ can be equivalently represented by the violation process $h_t(\alpha)$, which serves as the \textit{frequency} variable, as well as the cumulative violation on the $i-$ day of VaR violation, which we denote by $(H_i(\alpha))$, and referred to as the \textit{severity} variable. We call this approach frequency-severity, or ``two-part".\footnote{The term two-part is first introduced in the health econometrics literature \citep{cragg1971some, mullahy1998much} when dealing with health care cost, in case such care occurs. It is later introduced to the credit risk literature \citep{bruche2010recovery} when dealing with the probability of default and loss-given-default, in case of default. } This explicit separation of the VaR and ES components is also used in non-PIT based, but elicitability-based joint (VaR, ES) tests \citep{nolde2017elicitability,barendsebacktesting}. 

\subsection{From violation indicators to durations}

\begin{figure}[H]
\begin{center}
 \begin{tikzpicture}
\draw[gray, thick, ->] (0,0) -- (12,0);
\filldraw[black] (2,0) circle (2pt) node[anchor=west] {};
\filldraw[black] (5,0) circle (2pt) node[anchor=west] {};
\filldraw[black] (10,0) circle (2pt) node[anchor=west] {};
  \node (A) at (3.5,-0.4) {$d_i$};
    \node (B) at (7.5,-0.4) {$d_{i+1}$};
        \node (C) at (5,0.3) {$H_{i}$};
  \node (D) at (10,0.3) {$H_{i+1}$};
\end{tikzpicture}
 \end{center}
 \caption{Illustration of timeline of the different events. Each violation is represented by a dot symbol on the time axis. Each time a new violation occurs (say the $i-$th), one new duration ($d_i$) and one new severity $h_i$ becomes available.}
\label{fig:durations}
\end{figure}
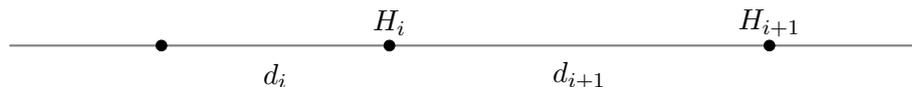

Rather than analyzing the frequency directly, we transform it into a sequence of durations as follows: $d_1 \geq $ represents the time elapsed until the first violation.\footnote{If the hit sequence starts with a violation, i.e., if $h_1>0$, we assume that $d_1=1$. In this case, our definition of $d_1$ is slightly different from \cite{christoffersen1998evaluating}, \cite{ pelletier2016geometric}, or \cite{candelon2011backtesting} who disregard the first value $d_1=1$ if the hit sequence starts with a violation. Since the probability of starting with a violation is quite small, the two definitions often lead to the same set of durations.} For $i \geq 2$, $d_i$ is the time elapsed between the $(i-1)$-th and the $i$-th violation. Figure \ref{fig:durations}  illustrates the timeline of the violations and their corresponding durations. Because the sequence of durations is in a one-to-one relationship with the sequence of violations, from now on, we will focus on the pair $(d_i, H_i)$. 

Duration-based backtests offer several advantages. First, as highlighted in the VaR literature \citep{christoffersen2004backtesting,Haas2006,candelon2011backtesting,pelletier2016geometric}, they may provide higher power compared to tests based on violations. Second, \cite{pelletier2016geometric} and \cite{du2017backtesting} argue that independence tests based on correlations can only detect linear serial dependence. In this respect, durations offer a nonlinear alternative to characterize the serial dependence of the violation process. Third, when backtesting ES, the number of observed severities is much smaller than the number of indicators for violations or no-violations. This asynchronous issue makes it difficult to analyze the interdependence between the frequency and severity of the violations. The introduction of durations solves this problem since each new VaR violation leads to a new duration and a new severity.

\medskip

The following property expresses the properties of the violation process $(h_t)$ equivalently in terms of the duration process.

\begin{Property}
    A sequence $(h_t, t=1,....)$ of Bernoulli variables is i.i.d. with probability $\alpha$, if and only if the associated duration sequence is i.i.d. with geometric distribution $\mathcal{G}eom(\alpha)$ with probability $\alpha$ and domain $\mathbb{N}^*$. The hit sequence $(h_t)$ is independent from the severities $(H_i)$ if and only if the duration sequence $(d_i)$ is independent of the severities $(H_i)$. 
\end{Property}

In this property, both the $i.i.d.$ and the geometric distribution assumptions are necessary. Indeed, without the geometric distribution constraint, the independence between the successive durations is also satisfied when the violation sequence $(h_t(\alpha))$ follows a two-state Markov chain studied in \cite{christoffersen1998evaluating}.
Similarly, without the independence assumption, the marginal $\mathcal{G}eom(\alpha)$ distribution alone for the durations does not imply the serial independence between the violation sequences either. Indeed, the sequence of durations can have both $\mathcal{G}eom(\alpha)$ marginal distribution and non-geometric conditional distribution \citep{gourieroux2019negative}. To our knowledge, the literature on duration-based VaR test has only focused on the test of the geometric distribution. 

\medskip

Consequently, the null hypothesis of ES validity we will backtest is as follows:
\vspace{0.1em}
\begin{hypothesis}
    $\mathcal{H}_0:$ the sequences $(d_i)$ and $(H_i)$ follow a $\mathcal{G}eom(\alpha)$ and $\mathcal{U}([0,1])$ distribution, respectively, and the pairs  $(H_i, d_{i+1})$, $(H_i, H_{i+1})$,  and $(d_i,d_{i+1})$  are mutually independent. 
\end{hypothesis}
\noindent This null hypothesis involves the whole left tail of the distribution of the PIT sequence $(u_t)$, as well as the joint tail, when $t$ varies. This is similar to the null hypothesis of \cite{berkowitz2001testing} and \cite{gordy2020spectral}, and is much stronger than standard mean-variance type moment conditions tested in the literature. It is also possible to add additional independence conditions between, say, $H_i$ and $H_{i+2}$, but since there are much less violations than there are trading days, such pairs are not necessarily observable. Thus they have not been included. 

\medskip

Two remarks should be made regarding the censored durations. Firstly, even though the first duration is left-truncated since it counts the time until the first violation, it also follows the geometric distribution, thanks to the lack of memory property of the geometric distribution. That is, given that a geometric variable $D$ is strictly larger than, say, $k \geq 1$, the truncated variable $D-k$ still follows the same geometric distribution. This means $d_1$ can be dealt with indiscriminately along with other uncensored durations.\footnote{This is helpful because previous studies recommend to remove $d_1$ from their analysis, which leads to a positive probability that the number of durations is too small and the test statistic cannot be computed. See for instance, Tables 1 and 2 (pages 324-326) in \cite{candelon2011backtesting}.} Secondly, after the last violation of the observation period, the next duration will be (right) censored. This censored duration does not follow the geometric distribution and is thus disregarded in our backtest, following \cite{candelon2011backtesting}.

\subsection{Characterizing joint distribution using (bivariate) orthogonal polynomials}

The test of the null $\mathcal{H}_0$ can be conveniently conducted using moment conditions involving orthogonal polynomials. While orthogonal polynomials have been used in finance for likelihood expansion \citep{ait2002maximum}, option pricing \citep{corrado1996skewness,xiu2014hermite}, and distribution testing \citep{bontemps2005testing, Bontemps_Meddahi_2012,candelon2011backtesting}, these works all focus on a univariate distribution, and have not been applied yet in a joint or dynamic setting such as ES backtesting. We are only aware of \cite{candelon2011backtesting} who use Meixner orthogonal polynomials associated to the geometric distribution to test marginal distributions of the duration in a VaR backtesting context.\footnote{Recently, (orthogonal) polynomials has also been used by \cite{horvath2022consistent} and \cite{Du2023} as regressors to approximate some univariate functions. Thus, in their approach, the orthogonal polynomial basis is not directly used for testing purpose and can be equivalently replaced by any polynomial basis such as the canonical polynomial basis $1, X, X^2, X^3,\dots$.}

\medskip

Let us denote by $(P_k(\cdot, \alpha))_{k \in \mathbb{N}}$ the sequence of Meixner orthogonal polynomials associated to the geometric $\mathcal{G}eom(\alpha)$ distribution, such that $P_k(\cdot, \alpha))$ is of degree $k$. Its first two terms are $P_0(X, \alpha)=1$ and $P_1(X, \alpha)=\frac{1-\alpha X}{\sqrt{1-\alpha}}$, and we refer to Appendix \ref{appendix:Meixner_polynomials} for its subsequent terms. By Theorem 1 of \cite{cameron1993tests}, any random variable $X$ follows the $\mathcal{G}eom(\alpha)$ distribution, if and only if:
\begin{equation}
    \label{alwaysorthogonal}
    \mathbb{E}[P_j(X, \alpha)]= 0, \qquad \forall j\geq 1.
\end{equation} 
The polynomials are called "orthogonal" (or orthonormal) because:
\begin{equation}
    \label{ortho1}
    \mathbb{E}[P_k(X, \alpha) P_j(X, \alpha)]= \mathbbm{1}(k=j), \qquad \forall k, j \geq 1
\end{equation}
where $\mathbbm{1}(.)$ denotes the indicator function. This condition will not be directly tested since \eqref{alwaysorthogonal} is already a characterization of the marginal geometric $\mathcal{G}eom(\alpha)$ distribution assumption. Nevertheless, Eq. \eqref{ortho1}, along with Eq. \eqref{alwaysorthogonal}, ensure that for $j\neq j'$, $P_j(X, \alpha)$ and $P_{j'}(X, \alpha)$ are uncorrelated. It will be seen later that this lack of correlation simplifies the distribution of our test statistics, by rendering the variance-covariance of the vector of sample moments an identity matrix. 

\medskip

Similarly, we denote by $(Q_k(\cdot))_{k \in \mathbb{N}}$ the sequence of orthogonal Legendre polynomials associated to the $\mathcal{U}([0,1])$ distribution. Its first two terms are $Q_0(Y)=1$ and $Q_1(Y)=\sqrt{3}(2Y-1)$, and its subsequent terms are given in Appendix \ref{appendix:Legendre_polynomials}. Then, a random variable $Y$ follows a $\mathcal{U}([0,1])$ distribution, if and only if:
\begin{equation}
    \label{orthogonal2}
    \mathbb{E}[Q_j(Y)]= 0, \qquad \forall j \geq 1.
\end{equation}
Moreover, in this case, we have the following orthogonality property: 
\begin{equation}
    \label{ortho2}
    \mathbb{E}[Q_k(Y) Q_j(Y)]= \mathbbm{1}(k=j), \qquad \forall k, j \geq 1.
\end{equation}
Similar as Eq. \eqref{ortho1}, Eq. \eqref{ortho2} will not be tested since Eq. \eqref{orthogonal2} already completely characterizes the uniform distribution on $[0,1]$.

\medskip

Such orthogonal polynomials can be extended to the bivariate case and give rise to a characterization of the independence between two variables. By Theorem 2 of \cite{cameron1993tests}, under the assumption that $X$ and $Y$ have  $\mathcal{G}eom(\alpha)$ and $\mathcal{U}([0,1])$ marginal distributions, respectively, are independent, if and only if:
\begin{equation}
    \label{crosscorrelation}
    \mathbb{E}[P_k(X, \alpha)Q_j(Y)]=0, \qquad \forall k, j \geq 1. 
\end{equation}
Similarly, if $X_1$ and $X_2$ both follow $\mathcal{G}eom(\alpha)$ distribution, then they are independent as soon as:
\begin{equation}
    \label{crosscorrelation2}
    \mathbb{E}[P_k(X_1, \alpha)P_j(X_2, \alpha)]=0, \qquad \forall k, j \geq  1.
\end{equation}
If both $Y_1$ and $Y_2$ follow $\mathcal{U}([0,1])$ distribution, then they are independent, if and only if:
\begin{equation}
    \label{crosscorrelation3}
    \mathbb{E}[Q_k(Y_1)Q_j(Y_2)]=0, \qquad \forall k, j \geq  1. 
\end{equation}

\subsection{The moment conditions to test} 
\label{moment_conditions}

Because \eqref{alwaysorthogonal}-\eqref{crosscorrelation3} each involves an infinity of equations, in practice, we have to limit ourselves to the first few conditions. Thus, the set of orthogonal conditions to test is:
\begin{align}
\mathbb{E}[Q_j(H_i)]&= 0, \qquad \forall j=1,...,K_1, \label{marginal1} \\
\mathbb{E}[P_j(d_i, \alpha)]&= 0, \qquad \forall j=1,...,K_2, \label{marginal2}\\
\mathbb{E}[P_k(d_i, \alpha)P_j(d_{i+1}, \alpha)]&=0, \qquad \forall k, j \geq 1, k+ j \leq K_3 ,\label{joint1} \\
\mathbb{E}[Q_k(H_{i+1})Q_j(H_i)]&=0, \qquad \forall  k, j \geq 1, k+ j  \leq K_4, \label{joint2}\\ 
\mathbb{E}[P_k(d_i, \alpha)Q_j(H_i)]&=0, \qquad \forall k, j \geq 1, k+j \leq K_5, \label{joint3}\\
\mathbb{E}[P_k(d_{i+1}, \alpha)Q_j(H_i)]&=0, \qquad \forall  k, j \geq 1, k+ j \leq K_6,\label{joint4} 
\end{align}
where integers $K_1, K_2$ re both greater than or equal to 1, while $K_3,...,K_6$ are non smaller than 2. For expository purpose, in the following we set $K_1=K_2=K$, and $K_3=K_4=K_5=K_6=K'$. Thus in total, we have $2K+2K'(K'-1)$ orthogonality conditions. 

\medskip

The above conditions are non-parametric (\textit{i.e.} model-free), since $i)$ they do not make assumptions on the DGP (as opposed to likelihood ratio tests); $ii)$ the number of equations to test can be made arbitrarily large, by selecting large values for $K$ and $K'$. In this respect, this approach shares some spirits with the spectral test of \cite{gordy2020spectral}. In some sense, both approaches can be regarded as extension of \cite{du2017backtesting}. \cite{gordy2020spectral} test (non orthogonal) moment conditions satisfied by a potentially large number of spectral transformations of the PIT, such as the cumulative violation. Their test statistic, however, $i)$ is not tailored to ES in the sense that most of the proposed spectral transformations are not directly related to ES; $ii)$ is (as a consequence) more complicated and could suffer from the curse of dimensionality, due to the need to estimate a high-dimensional covariance matrix. We, instead, focus on the single spectral transformation $(H_t(\alpha))$, and test moment conditions that are $i)$ directly motivated by ES, $ii)$ exhaustive, since \eqref{alwaysorthogonal} and \eqref{orthogonal2}-\eqref{crosscorrelation3} characterize the marginal and joint distributions, and $iii)$ simpler, thanks to the orthogonality property \eqref{ortho1}.

\medskip

Let us now interpret these conditions. The condition $\mathbb{E}[Q_j(H_i)]= 0$ for $j=1,...,K$ in Eq. \eqref{marginal1} is written on the marginal distribution of the severity, and corresponds to a UC test of ES. In particular, if $K=1$, then since $Q_1(Y)=\sqrt{3}(2Y-1)$, this condition becomes $\mathbb{E}[H_i]=\frac{1}{2}$. Thus we recover the UC test of \cite{du2017backtesting}. The motivation of testing this condition for higher moments $j>1$ is to increase the power of this UC test against DGP's such as those discussed in Section \ref{limitations}. For instance, if we take $K=4$, then the first four moment conditions combined allows to check the specification of the mean, variance, skewness and kurtosis of the severity variable. 

\medskip

Similarly, the condition $\mathbb{E}[P_j(d_i, \alpha)]= 0$ for $j=1,...,K$ (Eq. \eqref{marginal2}) is a duration-based UC test of the VaR, that focuses on the marginal distribution of the duration. 
If $K=1$, then the condition $\mathbb{E}[P_1(d_i,\alpha)]= 0$ reduces to $\mathbb{E}(d_i)=1/\alpha$, meaning that the expectation of the duration between two VaR violations should be equal to $1/\alpha$ if the bank's dynamic model is well specified. This corresponds to the UC duration-based VaR backtest proposed by \cite{christoffersen2004backtesting}, \citet{Haas2006}, \citet{candelon2011backtesting}, \cite{pelletier2016geometric}. If $K=4$ in Eq. \eqref{marginal2}, then this condition alone tests whether the marginal distribution of $d_i$ has the same mean, variance, skewness and kurtosis of as the geometric $\mathcal{G}eom(\alpha)$ distribution.  

\medskip

Conditions \eqref{joint1}-\eqref{joint4} characterize serial- or cross-sectional independence between the duration and violation variables. More precisely, the condition $\mathbb{E}[P_k(d_i, \alpha)P_j(d_{i+1}, \alpha)]=0$ tests the serial independence for the sequence of durations $(d_i)$. For instance, if $j=k=1$, then \eqref{joint1} becomes a condition of zero autocorrelation:
\begin{equation*}
    \mathbb{E}[(1-\alpha d_i)(1-\alpha d_{i+1})]=0.
\end{equation*}
This condition is for instance tested in the exponential autoregressive conditional duration (EACD) test of \cite{christoffersen2004backtesting}, which examines whether the coefficient in front of $d_i$ is zero in the regression of $d_{i+1}$ against $d_i$. If $(j,k)=(1,2)$ or $(j,k)=(2,1)$, then we get the analog of co-skewness \citep{friend1980co} of $d_i$ and $d_{i+1}$. If $(j,k)=(1,3), (2,2)$ or $(3,1)$, then we get the analog of co-kurtosis \citep{guidolin2008international} between $d_i$ and $d_{i+1}$. The introduction of higher order cross moments has a similar spirit as the recent proposal of \cite{Du2023}, who introduce nonlinear correlations to replace linear ones, in order to increase the power of the backtest of \cite{du2017backtesting}.

\medskip

Similarly, the condition $\mathbb{E}[Q_k(H_{i+1})Q_j(H_i)]=0$ tests serial independence for sequence $(H_i)$. The condition $\mathbb{E}[P_k(d_i, \alpha)Q_j(H_i)]=0$ tests whether in case of a violation, the time elapsed since the last violation has an effect on the severity of the current violation. Finally, the condition $\mathbb{E}[P_k(d_{i+1}, \alpha)Q_j(H_i)]=0$ tests whether the sequence $(H_i)$ Granger-causes the sequence $(d_i)$, that is, whether a particularly severe violation increases the likelihood of future violations. To our knowledge, such cross moments between the duration (or equivalently violation) sequence  and the severity have never been explored in the literature. In some sense, when testing the serial independence of each one of the two sequences (say, $(d_i)$), the lagged value of other sequence (say $H_{i-1}$) can be regarded as a covariate. Indeed, as \cite{gaglianone2011evaluating} put it, one weakness of many VaR tests is that by restricting to the duration/violation process only, too much information is sacrificed. Some attempts have been made to include observable covariates \citep{engle2004caviar, gaglianone2011evaluating, pelletier2016geometric, bayerregression} in order to increase the power, but often at the cost of rendering the tests more complicated. Our approach can be seen as a trade-off, in that it includes more information, but at the same time, it will become clear in the next section that its test statistics enjoys a highly tractable asymptotic $\chi^2$ distribution.  

\medskip

Furthermore, conditions \eqref{marginal1}-\eqref{joint4} can then be combined to construct various subtests of the UC or CC assumptions on the VaR, the ES or the pair (VaR, ES). These tests can be particularly useful for identifying the source of specification problems in an internal model. Here are a few examples:
\begin{enumerate}[$a)$]
    \item The conditions $\mathbb{E}[Q_j(H_i)]= 0$ and $\mathbb{E}[P_j(d_i,\alpha)]= 0$ form a UC test of the pair $(VaR, ES)$. 
    
    \item The conditions $\mathbb{E}[P_j(d_i,\alpha)]= 0$ and $\mathbb{E}[P_k(d_i, \alpha)P_j(d_{i+1}, \alpha)]=0$ can be viewed as a duration-based CC test of the VaR. It is the counterpart of the CC test in \cite{candelon2011backtesting, pelletier2016geometric}. Notably, condition \eqref{joint1}, which tests the independence of the durations, replaces the moment condition (16) of \cite{candelon2011backtesting}. The authors propose to use $\mathbb{E}[P_j(X,\beta)]$ with $\beta$ potentially different from the true level $\alpha$, as the IND condition.  Essentially, their alternative hypothesis still assumes the geometric distribution. Conversely, our test accommodates any alternative hypothesis, offering greater flexibility
    
    \item The conditions $\mathbb{E}[P_j(d_i,\alpha)]= 0$, $\mathbb{E}[P_k(d_i, \alpha)P_j(d_{i+1}, \alpha)]=0$, with $\mathbb{E}[P_k(d_{i+1}, \alpha)Q_j(H_i)]=0$ form another CC test of the VaR, but is an improvement of the previous test $b)$, since the inclusion of additional information on past violation severities makes it easier to detect potential mis-specifications of the VaR.
   
    \item Finally, the conditions $\mathbb{E}[Q_j(H_i)]= 0$ and $\mathbb{E}[P_j(d_i,\alpha)]= 0$, augmented with any nonempty subset of the joint moment conditions \eqref{joint2}-\eqref{joint4}, is a CC test of the pair $(VaR, ES)$. For instance, the joint conditions $\mathbb{E}[Q_j(H_i)]= 0$, $\mathbb{E}[P_j(d_i,\alpha)]= 0$, and $\mathbb{E}[Q_k(H_{i+1})Q_j(H_i)]=0$ form a counterpart of the CC test of \cite{du2017backtesting} and \cite{Du2023}.

\end{enumerate}

\subsection{The test statistic} 
An important advantage of orthogonal moment conditions is that under the null hypothesis, any univariate or bivariate polynomial taken from Eq. \eqref{marginal1}-\eqref{joint4} has zero mean and unit variance, and they are mutually uncorrelated. This suggests a Wald test. More precisely, for each of the $2K+ 2K'(K'-1)$ moment conditions, we replace the theoretical expectation on the left hand side by its empirical counterpart, and get a vector of dimension $2K+ 2K'(K'-1)$. Then the asymptotic variance covariance matrix of $V$ is simply the identity matrix. By central limit theorem, we deduce that when the number of violations $n$ increases to infinity:
\begin{equation}
    \label{bigtest}
    n V' V \xrightarrow{d} \chi^2(2K+ 2K(K'-1)),
\end{equation}
where $\chi^2(\nu)$ is the chi-squared distribution with $\nu$ degrees of freedom. Thus, at the test level $\kappa$, we reject the null hypothesis $\mathcal{H}_0$, if and only if the value of $n V'V$ is larger than the $(1-\kappa)$ quantile of the $\chi^2(2K+ 2K'(K'-1))$ distribution.  

\medskip

It is also possible to test only a subset of the moment conditions. Then the resulting test statistic is still asymptotically $\chi^2$, whose degree of freedom is equal to the number of conditions tested. Thus our testing approach also facilitates sub-tests, such as those mentioned at the end of Section \ref{moment_conditions}. This could be useful, for instance, if the global test (Eq. \ref{bigtest}) is rejected, the modeller might want to conduct different UC, IND and CC tests to understand which component(s) of the internal model is mis-specified. Such subtests have an advantage over two major existing VaR and ES backtesting approaches. At one extreme, many papers test UC and IND assumptions \textit{separately}, without conducting a \textit{joint} test. In other words, under these frameworks, a model is deemed acceptable, if and only if it passes both the UC and the CC tests. This creates a notoriously difficult, multiple testing problem \citep{vovk2022admissible}, since the probability that a model passes both tests might not be tractable. \cite{banulescu2021backtesting} propose to use Bonferroni correction, but this correction only provides a bound for the rejection rate. Our test, on the other hand, does not suffer from this difficulty, since the asymptotic distribution of any subtest is tractable. At the other end of the spectrum, some joint VaR-ES backtests, such as \cite{bayerregression}, are fundamentally difficult to separate into UC and IND conditions. In this sense, our test, which lies in between, is particularly interesting, since we can either split, or combine the different UC and IND conditions.  

\section{Monte-Carlo simulations}

In this section, we investigate the finite sample properties of our test by simulation. Specifically, we check that the empirical size of our test is close to its nominal level, and compare its power with existing backtests. Throughout this section, the tests are conducted at two ES coverage levels $\alpha = \{0.01, 0.05\}$, and the empirical rejection rates are calculated based on $1000$ replications, for a nominal size of $5\%$.

\subsection{Size of the test}

To evaluate the size of our test, we consider two data generating processes (DGP) that satisfy the null. The first DGP specifies the durations and severities directly. We simulate an i.i.d. sequence of durations $d_i$ from the $\mathcal{G}eom(\alpha)$ distribution, and an independent, i.i.d. sequence of severities $H_i$ from the $\mathcal{U}([0,1])$ distribution. The second DGP is a AR(1)-GARCH(1,1) model with student innovations. More precisely, we simulate minus returns from:
\begin{align}
    y_t &= \delta_0 + \delta_1 y_{t-1} + \epsilon_t, \label{garch1}\\
    \epsilon_t &= \sigma_t \eta_t, \quad \eta_t \sim t(\nu), \label{garch2} \\
    \sigma^2_t &= \gamma_0 + \gamma_1 \epsilon^2_{t-1} + \gamma_2 \sigma^2_{t-1}. \label{garch3}
\end{align}
Under this DGP, the true ES at coverage level $\alpha$ is defined as:
\begin{equation}
    ES_t(\alpha) = \delta_0 + \delta_1 y_{t-1} + \sigma_t \tau(\alpha),
\end{equation}
where $\tau(\alpha) = \mathbb{E}(\eta_t | \eta_t \ge F^{-1}_{\nu}(\alpha))$, and $F^{-1}_{\nu}(\alpha)$ is the $\alpha$-quantile of the student distribution. The calibration of parameters is similar to the one used in \citet{du2017backtesting}, that is,
$(\delta_0, \delta_1, \gamma_0, \gamma_1, \gamma_2, \nu) = (0, 0.05, 0.05, 0.1, 0.85, 5)$.

\medskip

The test is performed for several sample sizes $T = \{250; 500; 1000; 2500; 500000\}$, as well as different choices of $K = \{1;2;3;4\}$ and $K' = \{2;3\}$.\footnote{The results obtained for other values of $K$ and $K'$ are available upon request.} For each replication under the first DGP, a total of $T\alpha$ durations $d_i$ and severities $H_i$ are simulated, whereas under the second DGP, we simulate $T$ returns $y_t, t=1,...,T$, which approximately corresponds to $\alpha T$ durations $d_i$ and severities $H_i$. Table \ref{tab:size_alpha_005_001} displays the empirical size of the proposed test. 

\begin{table}[!htbp]
  \centering
  \caption{Empirical size of $5\%$ asymptotic test}
  \resizebox{15cm}{!}{
	\renewcommand{\arraystretch}{1.6}
    \tiny
    \begin{threeparttable}
    \begin{tabular}{lcccccccc}
    \midrule
    \multicolumn{9}{c}{First DGP} \\
    \midrule
    \multicolumn{9}{c}{$\alpha$ = 0.01} \\
    \midrule
          & \multicolumn{4}{c}{$K'$ = 2}    & \multicolumn{4}{c}{$K'$ = 3} \\
    \midrule
    Sample size & \multicolumn{1}{c}{$K$ = 1} & \multicolumn{1}{c}{$K$ = 2} & \multicolumn{1}{c}{$K$ = 3} & \multicolumn{1}{c}{$K$ = 4} & \multicolumn{1}{c}{$K$ = 1} & \multicolumn{1}{c}{$K$ = 2} & \multicolumn{1}{c}{$K$ = 3} & \multicolumn{1}{c}{$K$ = 4} \\
    \midrule
    T = 250 & 0.089 & 0.088 & 0.087 & 0.089 & 0.111 & 0.101 & 0.093 & 0.102 \\
    T = 500 & 0.079 & 0.089 & 0.079 & 0.077 & 0.089 & 0.099 & 0.101 & 0.105 \\
    T = 1000 & 0.097 & 0.083 & 0.067 & 0.093 & 0.090 & 0.102 & 0.099 & 0.083 \\
    T = 2500 & 0.063 & 0.075 & 0.079 & 0.057 & 0.117 & 0.104 & 0.083 & 0.082 \\
    T = 500000 & 0.048 & 0.051 & 0.060 & 0.058 & 0.053 & 0.043 & 0.051 & 0.059 \\
    \midrule
    \multicolumn{9}{c}{$\alpha$ = 0.05} \\
    \midrule
          & \multicolumn{4}{c}{$K'$ = 2}    & \multicolumn{4}{c}{$K'$ = 3} \\
    \midrule
    Sample size & \multicolumn{1}{c}{$K$ = 1} & \multicolumn{1}{c}{$K$ = 2} & \multicolumn{1}{c}{$K$ = 3} & \multicolumn{1}{c}{$K$ = 4} & \multicolumn{1}{c}{$K$ = 1} & \multicolumn{1}{c}{$K$ = 2} & \multicolumn{1}{c}{$K$ = 3} & \multicolumn{1}{c}{$K$ = 4} \\
    \midrule
    T = 250 & 0.074 & 0.079 & 0.084 & 0.083 & 0.099 & 0.094 & 0.103 & 0.110 \\
    T = 500 & 0.077 & 0.080 & 0.060 & 0.064 & 0.097 & 0.102 & 0.073 & 0.077 \\
    T = 1000 & 0.077 & 0.069 & 0.065 & 0.064 & 0.096 & 0.088 & 0.087 & 0.068 \\
    T = 2500 & 0.061 & 0.068 & 0.074 & 0.044 & 0.076 & 0.088 & 0.084 & 0.081 \\
    T = 500000 & 0.056 & 0.048 & 0.053 & 0.044 & 0.061 & 0.055 & 0.062 & 0.061 \\
    \midrule
    \multicolumn{9}{c}{Second DGP} \\
    \midrule
    \multicolumn{9}{c}{$\alpha$ = 0.01} \\
    \midrule
          & \multicolumn{4}{c}{$K'$ = 2}    & \multicolumn{4}{c}{$K'$ = 3} \\
    \midrule
    Sample size & \multicolumn{1}{c}{$K$ = 1} & \multicolumn{1}{c}{$K$ = 2} & \multicolumn{1}{c}{$K$ = 3} & \multicolumn{1}{c}{$K$ = 4} & \multicolumn{1}{c}{$K$ = 1} & \multicolumn{1}{c}{$K$ = 2} & \multicolumn{1}{c}{$K$ = 3} & \multicolumn{1}{c}{$K$ = 4} \\
    \midrule
    T = 250 & 0.029 & 0.039 & 0.050 & 0.055 & 0.054 & 0.053 & 0.071 & 0.052 \\
    T = 500 & 0.057 & 0.049 & 0.056 & 0.038 & 0.068 & 0.063 & 0.062 & 0.062 \\
    T = 1000 & 0.079 & 0.077 & 0.073 & 0.054 & 0.072 & 0.076 & 0.085 & 0.063 \\
    T = 2500 & 0.069 & 0.076 & 0.059 & 0.072 & 0.079 & 0.097 & 0.074 & 0.091 \\
    T = 500000 & 0.047 & 0.050 & 0.041 & 0.059 & 0.048 & 0.062 & 0.066 & 0.055 \\
    \midrule
    \multicolumn{9}{c}{$\alpha$ = 0.05} \\
    \midrule
          & \multicolumn{4}{c}{$K'$ = 2}    & \multicolumn{4}{c}{$K'$ = 3} \\
    \midrule
    Sample size & \multicolumn{1}{c}{$K$ = 1} & \multicolumn{1}{c}{$K$ = 2} & \multicolumn{1}{c}{$K$ = 3} & \multicolumn{1}{c}{$K$ = 4} & \multicolumn{1}{c}{$K$ = 1} & \multicolumn{1}{c}{$K$ = 2} & \multicolumn{1}{c}{$K$ = 3} & \multicolumn{1}{c}{$K$ = 4} \\
    \midrule
    T = 250 & 0.062 & 0.081 & 0.070 & 0.081 & 0.064 & 0.076 & 0.077 & 0.076 \\
    T = 500 & 0.082 & 0.073 & 0.077 & 0.050 & 0.090 & 0.080 & 0.084 & 0.086 \\
    T = 1000 & 0.074 & 0.066 & 0.058 & 0.051 & 0.095 & 0.086 & 0.090 & 0.084 \\
    T = 2500 & 0.058 & 0.061 & 0.060 & 0.064 & 0.073 & 0.084 & 0.070 & 0.082 \\
    T = 500000 & 0.065 & 0.075 & 0.054 & 0.047 & 0.056 & 0.061 & 0.045 & 0.064 \\
    \midrule
    \end{tabular}
    	\begin{tablenotes}[para,flushleft]
		\tiny
		\noindent Note: The results are based on $1000$ replications. For each sample, we report the percentage of rejection at a $5\%$ level.
		\end{tablenotes}
	\end{threeparttable}
  \label{tab:size_alpha_005_001}
  }
\end{table}

\medskip
 
Overall, the empirical size of the test is close to the nominal $5\%$ level, except for very small sample size such as $T=250$, where the test could be oversized. The size distortion seems also increasing in $K'$ and $K$, which control together the $2K+ K'(K'-1)$ moment conditions tested. This could be explained by the central limit theorem (CLT) which underlies the asymptotic theory of the test. For a given sample size $T$, the closer the distribution of the i.i.d. variables to the normal distribution, the better the quality of the normal approximation. For large $j$, the distribution of $P_j(d_i, \alpha)$ and $Q_j(H_i)$, where $d_i$ (resp. $H_i$) denotes the $i-$th duration (resp. severity), are quite far from the normality, since they are introduced precisely to test higher order properties, let alone the discrete (resp. bounded) support of the distribution. As an illustration, we plot below the histogram of $Q_3(H_i)$ and $Q_4(H_i)$ for $i=1,\dots,1000$. Finally, we confirm that the test is consistent. Regardless of the choice of $K$ and $K'$, the empirical size converges to the nominal size as $n$ increases.

\begin{figure}[H]
    \centering
        \includegraphics[scale=0.8]{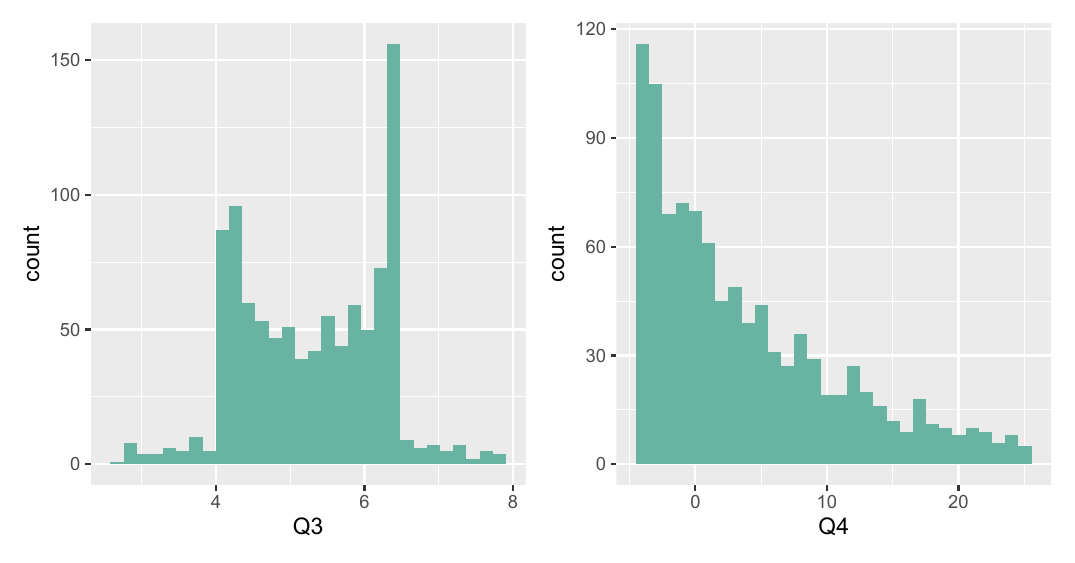}
    \label{fig:first}
          \caption{Histograms of $Q_3(H_i)$ (left panel) and $Q_4(H_i)$ (right panel) for 10000 severities $H_i$. Both distributions are highly non-normal. }
 \end{figure}

\medskip

Thus, we conclude that our backtest is valid, with only minor size distortion for small $T$. In the empirical application and the power experiments, this size distortion will be corrected by the Monte-Carlo simulation method of \cite{dufour2006monte}, following \cite{christoffersen2004backtesting}, \cite{berkowitz2011evaluating}, \cite{candelon2011backtesting}. This correction is quite straightforward, since the test statistic under the null hypothesis does not depend on any unknown parameter of the bank's model and is thus easy to sample from. 

\subsection{Power of the test}

Again, we consider two types of DGP's, one specified directly for $(d_i, H_i)$, whereas the other allows to generate daily minus returns.

\subsubsection{Alternative hypotheses written on durations and severities}

We consider 3 alternative DGPs denoted $A_1$,$A_2$, and $A_3$, respectively:  

\leftskip=1cm

\noindent $\mathbf{A_1}$: $d_i \sim \mathcal{G}eom(\alpha)$, $H_i \sim \mathcal{U}([0.2,0.8])$,  $t=1, \dots, T$, $(d_i)$ and $(H_i)$ mutually independent and i.i.d.

\noindent $\mathbf{A_2}$: $d_i-1 \sim NB((1-\alpha)/\alpha,0.5)$, where $NB$ denotes the negative binomial distribution, $(1-\alpha)/\alpha$ the number of successes, and $0.5$ the probability of success, $H_i \sim \mathcal{U}([0,1])$, $i=1, \dots, T$, $(d_i)$ and $(H_i)$ mutually independent and i.i.d.\footnote{Note that 1 is added to the simulated negative binomial variables to ensure that the support of $d_1$ is the set of positive integers.}

\noindent $\mathbf{A_3}$: $d_i-1 \sim NB((1-\alpha)/\alpha,0.5)$, $H_i \sim \mathcal{U}([0.2,0.8])$, $i=1, \dots, T$, $(d_i)$ and $(H_i)$ mutually independent and i.i.d.

\leftskip=0cm

\noindent Under $A_1$, only the marginal distribution of the severity is mis-specified,  but the distribution of the durations is correct, and the independence assumptions are valid. Since for the Legendre polynomials, $Q_1(H)=\sqrt{3}(2H-1)$, the mean of $H_i$ is well specified, implying that $\mathbb{E}[Q_1(H_i)]=0$ even under $A_1$. However, for any $j \geq 2$, we usually have $\mathbb{E}[Q_j(H_i)] \neq 0$ under $A_1$. Thus, if $K=1$ and $K' = 2$, then our test is expected to have low power against $A_1$, since it satisfies all the moment conditions. If we increase $K$ to at least 2, then some marginal moment conditions, such as $\mathbb{E}[Q_2(H_i)]=1$, are violated under $A_1$. Similarly, if we increase $K'$ to at least 4, then some joint moment conditions, such as $\mathbb{E}[Q_2(H_i)Q_2(H_{i+1})]=0$ are violated under $A_1$. Indeed, under $A_1$, the independence between $H_i$ and $H_{i+1}$ still holds, thus we still have:
\begin{equation*}
    \mathbb{E}[Q_2(H_i)Q_2(H_{i+1})]=\mathbb{E}[Q_2(H_i)]\mathbb{E}[Q_2(H_{i+1})]. 
\end{equation*}
But as explained above, under $A_1$, $\mathbb{E}[Q_2(H_i)]$ and $\mathbb{E}[Q_2(H_{i+1})]$ are no longer zero. Thus, the power of our test could be higher if $K \geq 2$, or if $K=1$, but $K'\geq 4$. Note, also, that $A_1$ is compatible with the motivational example discussed in Section \ref{limitations}, which satisfies both the UC and IND conditions tested by \cite{du2017backtesting}. 

\medskip

Similarly, under $A_2$, only the marginal distribution of the durations $d_i$ is mis-specified. Among the marginal moments, the expectation of $d_i$ is valid as under the null $\mathbb{E}(d_i)=1/\alpha$, but not the variance and other higher order moments. The condition $\mathbb{E}[P_1(d_i, \alpha)]=0$ is still satisfied under $A_2$, since the Meixner polynomials of order 1 is defined a $P_1(X, \alpha)=\frac{1-\alpha X}{\sqrt{1-\alpha}}$. However, the marginal conditions $\mathbb{E}[P_j(d_i, \alpha)]=0$ will not be satisfied for $j \geq 2$. Thus the power of our test is expected to be low against $A_2$ if $K=1$ and $K' = 2$, but will be higher if either $K\ge 1$, or $K' \ge 4$. 

\medskip

Finally, $A_3$ combines both mis-specifications of $A_1$ and $A_2$. Thus the impact of the values of $K$ and $K'$ under $A_3$ is expected to be similar as under $A_1$ and $A_2$, except that the power of the test is generally higher under $A_3$, because of the larger number of violated moment conditions. 

\begin{table}[!htbp]
  \centering
  \caption{Empirical size-corrected power of $5\%$ asymptotic test, $\alpha = 0.05$}
  \resizebox{15cm}{!}{
	\renewcommand{\arraystretch}{1.4}
    \tiny
    \begin{threeparttable}
    \begin{tabular}{lcccccccccc}
    \midrule
    \multicolumn{11}{c}{DGP: $A_1$} \\
    \midrule
          & \multicolumn{4}{c}{$K'$ = 2}    & \multicolumn{4}{c}{$K'$ = 3}    &       &  \\
    \midrule
    Sample size & \multicolumn{1}{c}{$K$ = 1} & \multicolumn{1}{c}{$K$ = 2} & \multicolumn{1}{c}{$K$ = 3} & \multicolumn{1}{c}{$K$ = 4} & \multicolumn{1}{c}{$K$ = 1} & \multicolumn{1}{c}{$K$ = 2} & \multicolumn{1}{c}{$K$ = 3} & \multicolumn{1}{c}{$K$ = 4} & \multicolumn{1}{l}{$BP_{ES}(5)$} & \multicolumn{1}{l}{$Z_C$} \\
    \midrule
    T = 250 & 0.011 & 0.049 & 0.038 & 0.048 & 0.021 & 0.027 & 0.030 & 0.036 &       &  \\
    T = 500 & 0.015 & 0.328 & 0.317 & 0.259 & 0.019 & 0.058 & 0.041 & 0.061 &       &  \\
    T = 1000 & 0.014 & 0.999 & 0.997 & 0.983 & 0.020 & 0.370 & 0.234 & 0.454 &       &  \\
    T = 2500 & 0.009 & 1.000 & 1.000 & 1.000 & 0.025 & 1.000 & 1.000 & 1.000 &       &  \\
    \midrule
    \multicolumn{11}{c}{DGP: $A_2$} \\
    \midrule
          & \multicolumn{4}{c}{$K'$ = 2}    & \multicolumn{4}{c}{$K'$ = 3}    &       &  \\
    \midrule
    Sample size & \multicolumn{1}{c}{$K$ = 1} & \multicolumn{1}{c}{$K$ = 2} & \multicolumn{1}{c}{$K$ = 3} & \multicolumn{1}{c}{$K$ = 4} & \multicolumn{1}{c}{$K$ = 1} & \multicolumn{1}{c}{$K$ = 2} & \multicolumn{1}{c}{$K$ = 3} & \multicolumn{1}{c}{$K$ = 4} & \multicolumn{1}{l}{$BP_{ES}(5)$} & \multicolumn{1}{l}{$Z_C$} \\
    \midrule
    T = 250 & 0.004 & 0.007 & 0.009 & 0.023 & 0.001 & 0.004 & 0.006 & 0.003 &       &  \\
    T = 500 & 0.002 & 0.012 & 0.266 & 0.760 & 0.000 & 0.002 & 0.009 & 0.039 &       &  \\
    T = 1000 & 0.004 & 0.076 & 0.999 & 1.000 & 0.001 & 0.002 & 0.158 & 0.996 &       &  \\
    T = 2500 & 0.003 & 1.000 & 1.000 & 1.000 & 0.000 & 0.501 & 1.000 & 1.000 &       &  \\
    \midrule
    \multicolumn{11}{c}{DGP: $A_3$} \\
    \midrule
          & \multicolumn{4}{c}{$K'$ = 2}    & \multicolumn{4}{c}{$K'$ = 3}    &       &  \\
    \midrule
    Sample size & \multicolumn{1}{c}{$K$ = 1} & \multicolumn{1}{c}{$K$ = 2} & \multicolumn{1}{c}{$K$ = 3} & \multicolumn{1}{c}{$K$ = 4} & \multicolumn{1}{c}{$K$ = 1} & \multicolumn{1}{c}{$K$ = 2} & \multicolumn{1}{c}{$K$ = 3} & \multicolumn{1}{c}{$K$ = 4} & \multicolumn{1}{l}{$BP_{ES}(5)$} & \multicolumn{1}{l}{$Z_C$} \\
    \midrule
    T = 250 & 0.000 & 0.000 & 0.006 & 0.048 & 0.000 & 0.000 & 0.000 & 0.000 &       &  \\
    T = 500 & 0.000 & 0.629 & 0.997 & 1.000 & 0.000 & 0.000 & 0.079 & 0.311 &       &  \\
    T = 1000 & 0.000 & 1.000 & 1.000 & 1.000 & 0.000 & 0.757 & 0.999 & 1.000 &       &  \\
    T = 2500 & 0.000 & 1.000 & 1.000 & 1.000 & 0.000 & 1.000 & 1.000 & 1.000 &       &  \\
    \midrule
    \multicolumn{11}{c}{DGP: $A_4$} \\
    \midrule
          & \multicolumn{4}{c}{$K'$ = 2}    & \multicolumn{4}{c}{$K'$ = 3}    &       &  \\
    \midrule
    Sample size & \multicolumn{1}{c}{$K$ = 1} & \multicolumn{1}{c}{$K$ = 2} & \multicolumn{1}{c}{$K$ = 3} & \multicolumn{1}{c}{$K$ = 4} & \multicolumn{1}{c}{$K$ = 1} & \multicolumn{1}{c}{$K$ = 2} & \multicolumn{1}{c}{$K$ = 3} & \multicolumn{1}{c}{$K$ = 4} & \multicolumn{1}{l}{$BP_{ES}(5)$} & \multicolumn{1}{l}{$Z_C$} \\
    \midrule
    T = 250 & 0.522 & 0.531 & 0.629 & 0.650 & 0.407 & 0.385 & 0.460 & 0.521 & 0.058 & 0.037 \\
    T = 500 & 0.755 & 0.887 & 0.890 & 0.938 & 0.611 & 0.763 & 0.769 & 0.845 & 0.048 & 0.034 \\
    T = 1000 & 0.989 & 0.999 & 0.996 & 0.998 & 0.930 & 0.967 & 0.985 & 0.991 & 0.052 & 0.044 \\
    T = 2500 & 1.000 & 1.000 & 1.000 & 1.000 & 1.000 & 1.000 & 1.000 & 1.000 & 0.054 & 0.042 \\
    \midrule
    \multicolumn{11}{c}{DGP: $A_5$} \\
    \midrule
          & \multicolumn{4}{c}{$K'$ = 2}    & \multicolumn{4}{c}{$K'$ = 3}    &       &  \\
    \midrule
    Sample size & \multicolumn{1}{c}{$K$ = 1} & \multicolumn{1}{c}{$K$ = 2} & \multicolumn{1}{c}{$K$ = 3} & \multicolumn{1}{c}{$K$ = 4} & \multicolumn{1}{c}{$K$ = 1} & \multicolumn{1}{c}{$K$ = 2} & \multicolumn{1}{c}{$K$ = 3} & \multicolumn{1}{c}{$K$ = 4} & \multicolumn{1}{l}{$BP_{ES}(5)$} & \multicolumn{1}{l}{$Z_C$} \\
    \midrule
    T = 250 & 0.124 & 0.114 & 0.124 & 0.107 & 0.121 & 0.117 & 0.137 & 0.096 & 0.122 & 0.061 \\
    T = 500 & 0.127 & 0.167 & 0.188 & 0.205 & 0.152 & 0.181 & 0.158 & 0.157 & 0.160 & 0.069 \\
    T = 1000 & 0.169 & 0.224 & 0.257 & 0.285 & 0.203 & 0.269 & 0.232 & 0.217 & 0.204 & 0.054 \\
    T = 2500 & 0.292 & 0.423 & 0.410 & 0.401 & 0.308 & 0.361 & 0.386 & 0.393 & 0.397 & 0.049 \\
    \midrule
    \end{tabular}%
      	\begin{tablenotes}[para,flushleft]
		\tiny
		\noindent Note: The results are based on $1000$ replications and are size-corrected. For each sample, we report the percentage of rejection at a $5\%$ level. The conditional test statistic of \citet{du2017backtesting} is denoted as $BP_{ES}(m)$, where $m$ is the degree of freedom, and while $Z_C$ corresponds to the conditional test statistic of \citet{acerbi2014back}.
		\end{tablenotes}
	\end{threeparttable}
  \label{tab:power_alpha_005}
  }
\end{table}

\medskip

The first three sub-tables in Table \ref{tab:power_alpha_005} report the size-corrected power of our test against $A_1, A_2$ and $A_3$, respectively.\footnote{The size-corrected power of a test is computed by comparing its statistic to the quantile, at the nominal size of $5\%$, of the test statistic measured under the null. Results for $\alpha = 0.01$ are reported in Table \ref{tab:power_alpha_001} in Appendix \ref{appendix:additional_tables}.} As expected, the power against $A_1$ is the lowest when $K = 1$ and $K' = 2$. Interestingly, the power of the test is the highest for $(K' = 2, K = 2)$, instead of for $(K' = 3, K = 2)$, or $(K'=3, K=4)$. The power (not reported) is also low when $(K' = 4, K = 2)$. This can be explained by the fact that the total number of moment conditions increases quadratically (resp. linearly) in $K'$ (resp. $K$). Thus, large $K$ and/or $K'$ could lead to too many moment conditions, among which only a small fraction are violated by the alternative hypothesis, hence potentially diminishing the power of the test. The conclusions we can draw from $A_2$ and $A_3$ are similar. However, the power is the highest for $(K' = 2, K = 4)$, as moment conditions related to higher moments of duration and severities are largely violated, unlike $A_1$. Moreover, the power decreases with $K'$, as too many moment conditions are tested, and only those related to the marginal distribution of $d_i$ or $H_i$, (i.e., Eqs. \eqref{marginal1} and \eqref{marginal2}), are violated.

\medskip

The comparison of the power for different values of $K$ and $K'$ suggests that in practice, their optimal choice depends on the PIT data at hand. Intuitively, if some low order moment conditions are violated, then small $K$ and $K'$ might suffice. On the other hand, if the underlying DGP differs from the null only moderately, for instance through some higher order polynomial moments only, then we need to increase $K$ and $K'$ o enhance the rejection rate of the test. In the literature, the selection of optimal number of moment conditions is first proposed by \cite{2006Towards}, and has been applied in the finance literature by \cite{escanciano2009automatic} and \cite{Du2023}. These methods are based on the idea that polynomial expansions can be used to approximate some (univariate) functions such as a pdf or a spectral density. This approach, however, cannot be directly applied in our framework, since our test involves several bivariate distributions. Thus, to choose $K$ and $K'$, we propose the following rule-of-thumb.
\begin{recommendation}[Choice of $K$ and $K'$]
\noindent
\begin{itemize}
    \item Start the test with $K = 1$ and $K' = 2$.
    \begin{itemize}
        \item If the null is rejected, then stop here as it implies that conditions on the expectation and/or serial/mutual independence of the duration and severity sequences are violated.  
        \item If the null is not rejected, then increase $K$ to check if conditions on higher order moments of durations and/or severities
        are violated. 
    \end{itemize}
    \item If the null is still not rejected for a large value of $K$, such as $K = 4$, then reset $K$ to 1 and 
    increase  $K'$ to check if conditions on higher order cross-moments are violated. 
    \item If the null is still not rejected for large values of $K$ and $K'$, such as $(K, K') = (4, 5)$, stop here as the power of the test usually starts to decrease in $K$ and $K'$ when they are sufficiently large. 
\end{itemize}
\end{recommendation}

Another way of adjusting the total number of moment conditions is to include only a subset of the equations among \eqref{marginal1}-\eqref{joint4}. For instance, at the end of Section \ref{moment_conditions}, we have listed four such combinations, with different interpretations as UC/CC test of either VaR, or the pair (VaR,ES). As an illustration, Table \ref{tab:power_alpha_005_CC_VaR_duration} in Appendix \ref{appendix:additional_tables} reports the power of our duration-based CC VaR subtest, based on \eqref{marginal2} and \eqref{joint1}.\footnote{Table \ref{tab:size_alpha_005_001_CC_VaR_duration} in Appendix \ref{appendix:additional_tables} reports the size of our duration-based CC VaR subtest.} Because these conditions are all satisfied by $A_1$, this test has lower power against $A_1$ and as expected, this is confirmed by the first sub-table of Table \ref{tab:power_alpha_005_CC_VaR_duration}, which shows that this subtest has a size-corrected power barely equal to the nominal size. On the other hand, the second and third sub-table of Table \ref{tab:power_alpha_005_CC_VaR_duration} confirm that this subtest is capable of detecting the mis-specification of the durations under $A_2$ and $A_3$. We can also compare the power of the global test with those of the subtests against $A_2$ and $A_3$. The global test has a larger size-corrected power under $A_3$, as many of the additional moment conditions included in the global test, but not in the subtest, are violated under $A_3$. On the other hand, the global test has a slightly lower power against $A_2$, since the additional moment conditions are not violated under $A_2$. 

\medskip

The above comparisons suggest that it could be useful to combine the results of both the global test, and some of the subtests, to gain insights on the different components of the model, regardless of the result of the global test. We suggest the following procedure:
\begin{recommendation}[Global test \textit{vs} subtests]
\label{recommendation2}
\noindent
\begin{itemize}
    
    \item If the global test rejects the null, but not a subtest, 
    then the conditions involved in the subtests are not responsible for the failure of the test. On the other hand, if both the global test and some subtests reject the null, then the components of the model tested by the subtest clearly needs to be improved.
     
    \item If the global test does not reject the null, it might still be interesting to conduct some of the subtests, since for certain alternative hypotheses, some subtests have higher power. In particular, if some subtests reject the null, then the components of the model involved in that subtest might still deserve improvement, even if the global test not rejecting the null. 
    
\end{itemize}
\end{recommendation}

Finally, qualitatively similar results are displayed in Tables \ref{tab:power_alpha_001} and \ref{tab:power_alpha_001_CC_VaR_duration} for $\alpha = 0.01$. Generally, we observe that \textit{ceteris paribus}, the power is lower for $\alpha = 0.01$ than for $\alpha = 0.05$, due to fewer VaR violations. 

\subsubsection{Alternative hypotheses written on the returns}

We now consider a GARCH-type null hypothesis against two alternative hypotheses, and use them to compare our test with two popular ES backtests, that are the IND test of \citet{du2017backtesting} and the test of \citet{acerbi2014back}. The former can be represented by the condition \eqref{inddu} in our framework, whereas the latter corresponds to condition \eqref{acerbi}. In this experiment, we simulate the returns using the AR(1)-GARCH(1,1) model defined in Eqs. \eqref{garch1}-\eqref{garch3}.

\medskip

\noindent The two alternative hypotheses we consider correspond to two mis-specified internal models, respectively denoted $A_4$ and $A_5$.

\leftskip=0.6cm

\noindent $\mathbf{A_4}$: A bank mis-specifies the distribution of the innovations $\eta_t$ in the AR(1)-GARCH(1) model, by assuming standard normal instead. For our test and the one of \citet{du2017backtesting}, the standard normal distribution is used to compute the conditional PIT, whereas for the test of \citet{acerbi2014back}, the ES is computed based on $\mu(\alpha) = \mathbb{E}(\eta_t | \eta_t \ge F^{-1}(\alpha))$, where $F^{-1}(\alpha)$ is the $\alpha$-quantile of the standard normal distribution.  

\noindent $\mathbf{A_5}$: Another bank uses wrong parameter values $(\delta_0, \delta_1, \gamma_0, \tilde{\gamma}_1, \tilde{\gamma}_2, v) = (0, 0.05, 0.05, 0.04, 0.91, 5)$ to compute the conditional variance. 

\leftskip=0cm

\noindent As we have explained in Section \ref{limitations}, the test of \citet{acerbi2014back} is expected to have no power against $A_4$. Similarly, since this alternative does not introduce autocorrelation for $(H_i)$, the IND backtest of \citet{du2017backtesting} has low power against it.\footnote{Their UC test, on the other hand, does have power against $A_4$. This test is omitted, to save space.} On the other hand, since the distributions of both the durations $d_i$ and severities $H_i$ deviate from their respective null, our test is expected to have a large power against this alternative. Under $A_5$, the durations and severities calculated by the bank can feature residual auto- and cross-correlation. Therefore, both our test and the IND backtest of \citet{du2017backtesting} might detect this autocorrelation and have power against $A_5$.

\medskip

The last two sub-tables in Tables \ref{tab:power_alpha_005} and \ref{tab:power_alpha_001} display the empirical size-corrected power of the different tests at coverage level $\alpha = 0.05$ and $\alpha = 0.01$, against $A_4$ and $A_5$. As expected, our test has a large power against $A_4$, compared to those of \citet{du2017backtesting} and \citet{acerbi2014back}. Against $A_5$, our test and the one of \citet{du2017backtesting} have similar power, whereas the test of \citet{acerbi2014back} has low power. Notice that for both alternative hypotheses, the power increases in $K$ as moments related to high order moments are violated. However, the power decreases with $K'$. These results are thus consistent with the rule-of-thumb that we suggest to select $K$ and $K'$. Similar results (not reported) are obtained for the other sub UC/CC tests presented in Section \ref{moment_conditions}. Results are available upon request. 

\section{Backtesting systemic risk measures}
The methodology introduced in this paper can also be applied to backtest many systemic risk measures, that are used by regulators to identify the institutions most vulnerable to a general market downturn. For instance, recently, \cite{banulescu2021backtesting} adapt the approach of \cite{du2017backtesting} to backtest Marginal Expected Shortfall (MES) of a firm's stock return $Y_{1,t}$ with respect to the market return $Y_{2,t}$. Just as the ES, which is defined indirectly through the VaR, the MES is defined indirectly through CoVaR \citep{Adrian_CoVaR}. At the coverage level $\alpha$ for the market return, the $CoVaR_{1t}(\beta,\alpha)$ at level $\beta$ for the firm at $t$ is defined by:
\begin{equation*}
    \mathbb{P}[Y_{1t} \leq - \text{CoVaR}_{1t}(\beta,\alpha)|Y_{2t} \leq -VaR_{2t}(\alpha) ]=\beta, \qquad \forall \beta \in [0,1],
\end{equation*}
where $VaR_{2t}$ is the VaR of the market return at date $t$. Then the MES of this firm at (market) level $\alpha$ at $t$ is an integral of the CoVaR:
\begin{equation}
    \label{mes}
    MES_t(\alpha)=\int_0^1  \text{CoVaR}_{1t}(\beta, \alpha) \mathrm{d} \beta.
\end{equation}
Similar to ES, the MES is difficult to backtest directly. Thus, \cite{banulescu2021backtesting} introduce the joint violation process,
\begin{equation}
    h_t(\alpha, \beta) = \mathbbm{1}(Y_{1t} \leq  CoVaR_{1t}(\beta, \alpha)) \mathbbm{1}(Y_{2t} \leq  VaR_{2t}(\alpha)),
\end{equation}
which is the analog of violation process defined in \eqref{violation}, and then integrate it to get the MES analog of the cumulative violation, called cumulative joint violation: 
\begin{equation}
    \label{MEScumulative}
    \tilde{H}_t(\alpha)=\int_0^1  h_t(\alpha, \beta) \mathrm{d} \beta.
\end{equation}
This latter can also be equivalently written through PIT:
\begin{equation}
    \label{mespit}
    \tilde{H}_t(\alpha)=u_{12t} \mathbbm{1}(u_{2t}\leq \alpha),
\end{equation}
where $u_{2t}$ is the conditional PIT of the market return, and $u_{12t}$ is the conditional PIT of the return of the firm, given that the market return violates its conditional VaR. 

\medskip

Then \cite{banulescu2021backtesting} use $\mathbb{E}[\tilde{H}_t(\alpha)]=\alpha/2$ for the UC backtest and its lack of autocorrelation for the CC test. They also show that the same methodology can  be applied to other systemic risk measures, such as the Systemic Expected Shortfall (SES) of \cite{acharya2017measuring}, the SRISK of \cite{acharya2012capital}, and the $\Delta$CoVar of \cite{Adrian_CoVaR}. 

\medskip

However, the downsides of these UC and IND tests are similar to those of \cite{du2017backtesting}, in that $\tilde{H}_t(\alpha)$ is also observed only when a VaR violation occurs for the market. To apply our methodology, it suffices to remark that the sequence of $\tilde{H}_t(\alpha)$ can also be equivalently represented by the sequence of durations $(d_i)$ between consecutive market VaR violations, as well as the severity variable $(\tilde{H}^{(1)}_i)$, which are equal to $u_{12t}$, on dates $t$ where market VaR violation occurs. If the model is correctly specified, then the durations are i.i.d. geometric distributed, whereas $u_{12t}$ is i.i.d., uniformly distributed and independent of the durations. As a consequence, the orthogonality conditions \eqref{marginal1}-\eqref{joint4} can be directly applied to these two sequences. 

\medskip

Similarly, the use of a potentially large number of moment conditions has the advantage of increasing the power of the backtest. Indeed, as \cite{fissler2023backtesting} put it, the test of \cite{banulescu2021backtesting} uses one-dimensional identification functions for $(VaR_{2t}(\alpha), CoVaR_{1t}(\beta, \alpha))$, which lacks power against certain mis-specified alternatives. The solution proposed by \cite{fissler2023backtesting}, however, is a joint backtest of VaR and MES, which does not allow to backtest MES alone. Our approach solves this issue, while also improving the power of the \cite{banulescu2021backtesting} test. 

\section{Empirical application}

\begin{figure}[!htbp]
    \centering
    \includegraphics[width=1\linewidth]{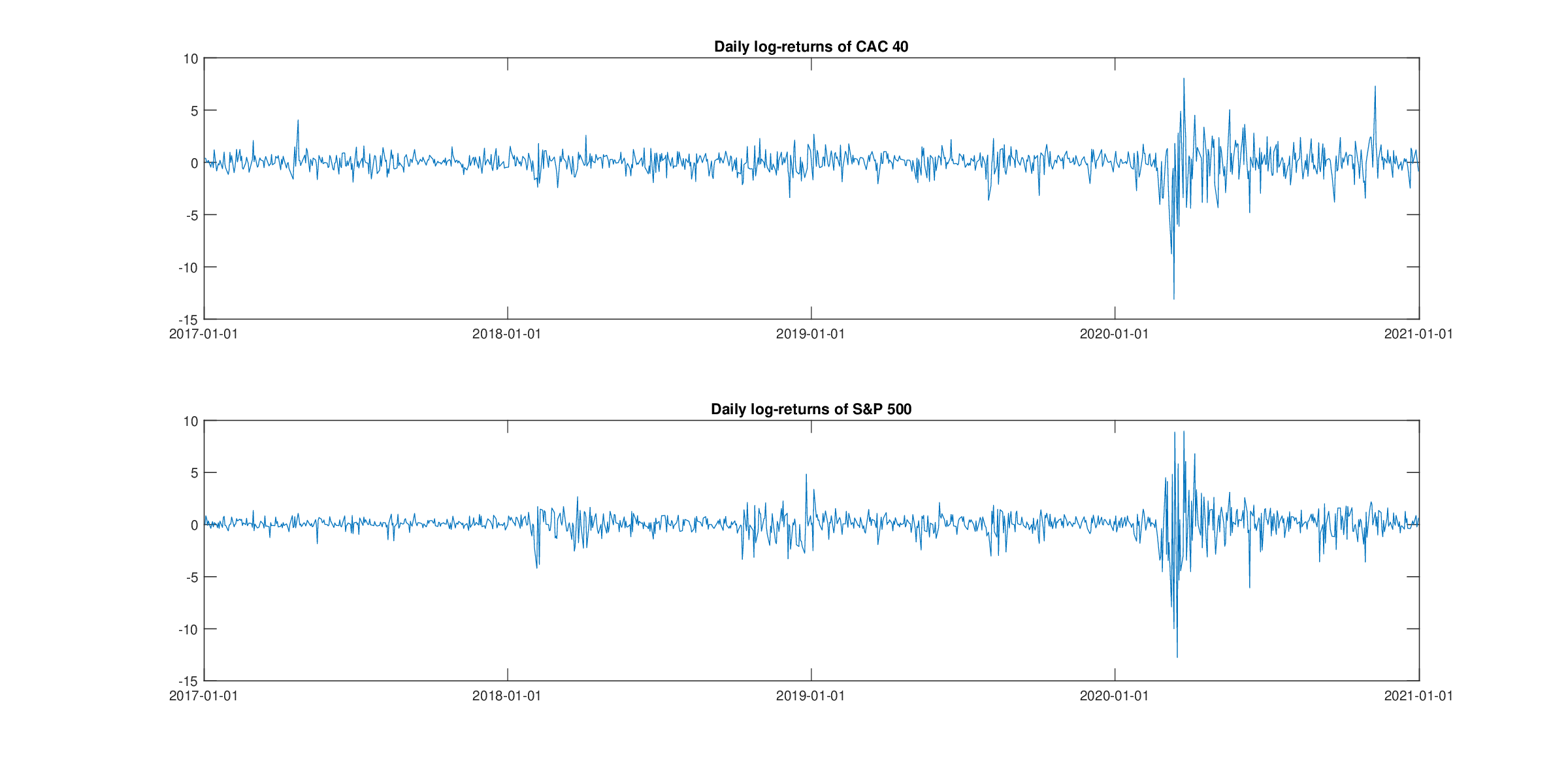}
    \caption{Plots of daily log-returns of CAC 40 (upper panel) and S\&P 500 (lower panel) for both the in-sample and out-of-sample periods.}
    \label{fig:returns}
\end{figure}

We apply our test to daily log-returns of CAC 40 and S\&P 500 indices over a period of four years from January 1, 2017 to December 31, 2020, totaling roughly 1000 trading days. The log-returns of both indices are displayed in Figure \ref{fig:returns}, and their basic summary statistics are reported in Table \ref{tab:summary_statistics}. The first three years of data serve as the training sample for the volatility model. Following \citet{bis2019}, we consider a one-year out-of-sample period between January 1, 2020 and December 31, 2020, which overlaps with the COVID crisis, when returns are more volatile, negatively skewed and leptokurtic. 

\begin{table}[!htbp]
  \centering
  \caption{In-sample and out-of-sample summary statistics for the two indices}
    \begin{tabular}{lccccc}
    \midrule
          & \multicolumn{2}{c}{In-sample (2017, 2018, 2019)} &       & \multicolumn{2}{c}{Out-of-sample (2020)} \\
    \midrule
          & \multicolumn{1}{l}{CAC40} & \multicolumn{1}{l}{S\&P500} &       & \multicolumn{1}{l}{CAC40} & \multicolumn{1}{l}{S\&P500} \\
    \midrule
    No. of obs. & 766 & 782 &       & 257 & 262 \\
    Mean  & 0.027 & 0.047 &       & -0.029 & 0.057 \\
    Median & 0.046 & 0.051 &       & 0.032 & 0.182 \\
    St. Dev. & 0.795 & 0.794 &       & 2.058 & 2.147 \\
    Skewness & -0.270 & -0.713 &       & -1.126 & -0.884 \\
    Exc. of Kurto. & 2.376 & 5.765 &       & 8.501 & 9.108 \\
    Maximum & 4.060 & 4.840 &       & 8.056 & 8.968 \\
    Minimum & -3.635 & -4.184 &       & -13.098 & -12.765 \\
    10 percentile & -0.936 & -0.691 &       & -1.922 & -1.826 \\
    5 percentile & -1.413 & -1.361 &       & -3.680 & -3.416 \\
    1 percentile & -2.096 & -2.710 &       & -6.106 & -7.682 \\
    \midrule
    \end{tabular}
  \label{tab:summary_statistics}
\end{table}

\newpage

\begin{figure}[!htbp]
    \centering
    \includegraphics[width=1\linewidth]{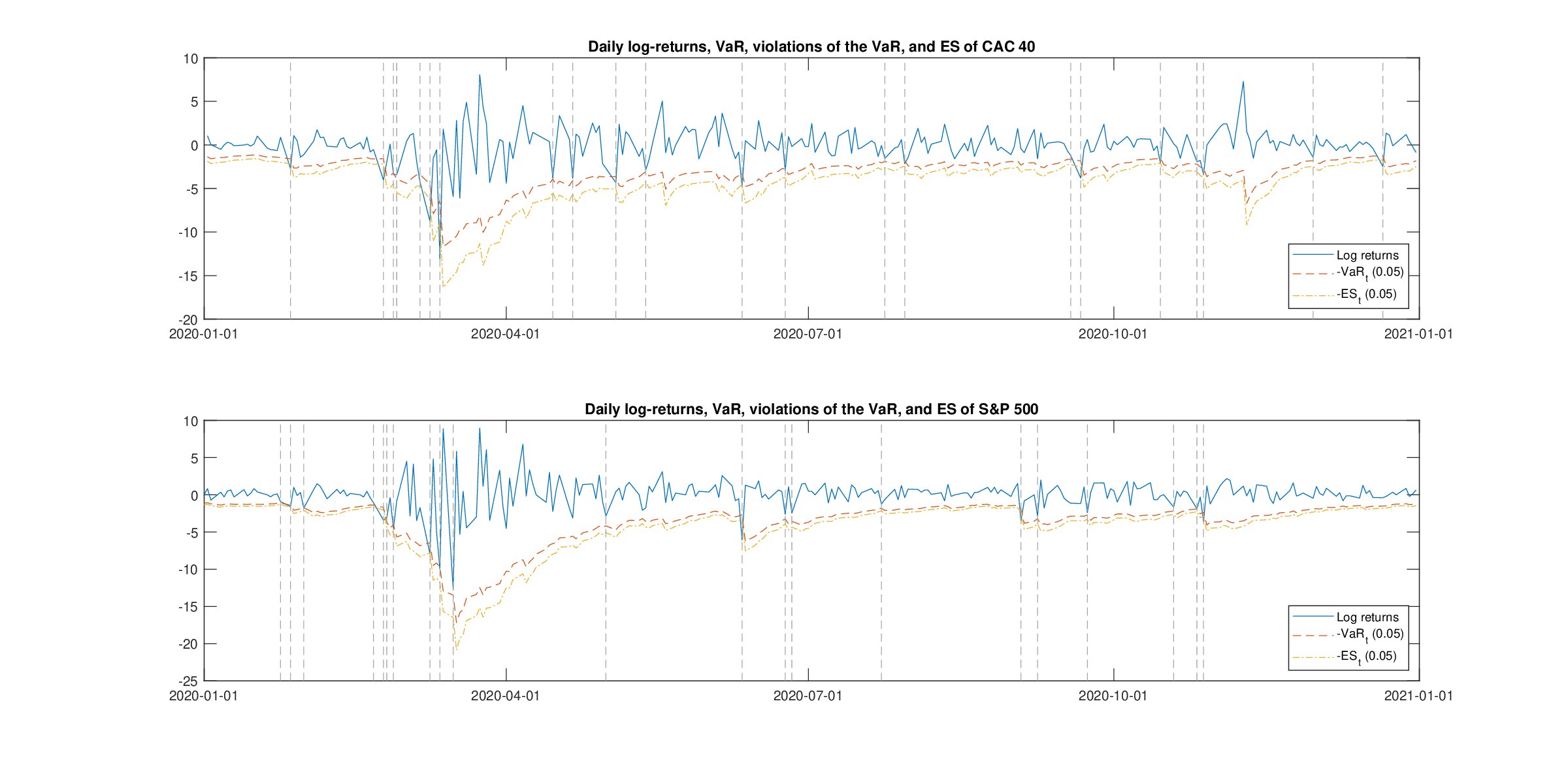}
    \caption{Plots of out-of-sample daily log-returns (blue full line), VaR (red dashed line), ES (orange dashed line), as well as days of VaR violations (horizon lines) for CAC40 (upper panel) and S\&P 500 (lower panel).}
    \label{fig:returns_VaR_ES}
\end{figure}

For each index, we use maximum likelihood to estimate a $t$-AR(1)-GARCH(1,1) model on the in-sample data (see Tables \ref{tab:estimated_model} for the parameter estimate), and perform Ljung-Box tests on the standardized residuals as well as their squares (see Table \ref{tab:LB_test} in Appendix \ref{appendix:additional_tables}) to validate the goodness-of-fit of the model. Using the fitted models, we forecast the VaR and ES at the $\alpha = 0.05$ coverage level during the out-of-sample period. Figure \ref{fig:returns_VaR_ES} plots the VaR and ES forecasts, and identifies the dates of the VaR violations in dotted lines. For both indices, we observe a cluster of violations at the beginning of the COVID crisis, suggesting potentially inadequate VaR (and hence ES) forecasts.

Table \ref{tab:pvalues_backtests_CAC40} and \ref{tab:pvalues_backtests_SP500} report the $p$-values of the global test and of the four subtests presented in Section \ref{moment_conditions} for CAC40 and S\&P500, respectively.
For the CAC 40, the null hypothesis $\mathcal{H}_0$ is rejected at the $5\%$ significance level for $K' = 2$ by the global test. To understand the cause of this rejection, let us resort to the subsets. First, CC VaR tests do not reject the null regardless of the values of $K$ and $K'$, while the other tests reject for all values of $K$ and $K'$. This suggests that the rejections are mainly due to the mis-specification of severity $H_i$. In other words, Eqs. \eqref{marginal2} and \eqref{joint1} are likely not violated. Moreover, as $p$-values of the CC VaR test are larger than those of the CC VaR duration-based test, we can also deduce that that moment conditions of Eq. \eqref{joint4} are likely not violated either.
Finally, the CC and UC tests of the pair (VaR, ES) reject the null for all values of $K$ and $K'$. These results highlight that moment conditions of Eq. \eqref{marginal1} are violated regardless of $K$, but also imply that condition \eqref{joint2} is violated, even for large values of $K'$. These results are consistent with our first conclusion that the rejection of the global test mainly comes from the severities rather than from the durations, and thus that the ES is mis-specified.
\begin{table}[!htbp]
  \centering
  \small 
  \caption{$p$-values for different backtests of the VaR and ES for CAC40}
  \begin{threeparttable}
    \begin{tabular}{lcccccc}
    \midrule
    \multicolumn{2}{c}{Test} & Global & CC (duration) & CC & CC & UC \\
    \midrule
    \multicolumn{2}{c}{Measure} & (VaR, ES) & VaR & VaR & (VaR, ES) & (VaR, ES) \\
    \midrule
    \multicolumn{2}{c}{Equations} & \eqref{marginal1}-\eqref{joint4} & \eqref{marginal2},\eqref{joint1} & \eqref{marginal2}, \eqref{joint1}, \eqref{joint4} & \eqref{marginal1}, \eqref{marginal2}, \eqref{joint2} & \eqref{marginal1}, \eqref{marginal2} \\
    \midrule
    \multicolumn{1}{c}{$K$} & \multicolumn{1}{c}{$K'$} & \multicolumn{5}{c}{} \\
    \midrule
    1     & 2     & \textbf{0.034} & 0.094 & 0.148 & \textbf{0.009} & \textbf{0.003} \\
    2     & 2     & \textbf{0.044} & 0.120 & 0.167 & \textbf{0.026} & \textbf{0.017} \\
    3     & 2     & \textbf{0.043} & 0.145 & 0.195 & \textbf{0.026} & \textbf{0.019} \\
    4     & 2     & \textbf{0.050} & 0.172 & 0.226 & \textbf{0.035} & \textbf{0.027} \\
    \midrule
    1     & 3     & 0.111 & 0.172 & 0.334 & \textbf{0.016} & \textbf{0.003} \\
    2     & 3     & 0.108 & 0.187 & 0.337 & \textbf{0.032} & \textbf{0.017} \\
    3     & 3     & 0.105 & 0.215 & 0.364 & \textbf{0.032} & \textbf{0.019} \\
    4     & 3     & 0.111 & 0.243 & 0.392 & \textbf{0.043} & \textbf{0.027} \\
    \midrule
    1     & 4     & 0.116 & 0.251 & 0.474 & \textbf{0.007} & \textbf{0.003} \\
    2     & 4     & 0.113 & 0.256 & 0.470 & \textbf{0.018} & \textbf{0.017} \\
    3     & 4     & 0.112 & 0.282 & 0.494 & \textbf{0.019} & \textbf{0.019} \\
    4     & 4     & 0.116 & 0.309 & 0.516 & \textbf{0.023} & \textbf{0.027} \\
    \midrule
    1     & 5     & 0.151 & 0.322 & 0.589 & \textbf{0.010} & \textbf{0.003} \\
    2     & 5     & 0.147 & 0.326 & 0.580 & \textbf{0.017} & \textbf{0.017} \\
    3     & 5     & 0.144 & 0.352 & 0.600 & \textbf{0.019} & \textbf{0.019} \\
    4     & 5     & 0.148 & 0.378 & 0.620 & \textbf{0.023} & \textbf{0.027} \\
    \midrule
    \end{tabular}%
    \begin{tablenotes}[para,flushleft]
		\footnotesize
		\noindent Note: This table displays the $p$-values of different backtests of the VaR and ES.  The $p$-values have been obtained based on the approach of \citet{dufour2006monte}, and those lower than $5\%$ are displayed in bold.
		\end{tablenotes}
	\end{threeparttable}
 \label{tab:pvalues_backtests_CAC40}
\end{table}%

\newpage

\begin{table}[!htbp]
  \centering
  \small 
  \caption{$p$-values for different backtests of the VaR and ES for S\&P 500}
  \begin{threeparttable}
    \begin{tabular}{lcccccc}
    \midrule
    \multicolumn{2}{c}{Test} & Global & CC (duration) & CC & CC & UC \\
    \midrule
    \multicolumn{2}{c}{Measure} & (VaR, ES) & VaR & VaR & (VaR, ES) & (VaR, ES) \\
    \midrule
    \multicolumn{2}{c}{Equations} & \eqref{marginal1}-\eqref{joint4} & \eqref{marginal2},\eqref{joint1} & \eqref{marginal2}, \eqref{joint1}, \eqref{joint4} & \eqref{marginal1}, \eqref{marginal2}, \eqref{joint2} & \eqref{marginal1}, \eqref{marginal2} \\
    \midrule
    \multicolumn{1}{c}{$K$} & \multicolumn{1}{c}{$K'$} & \multicolumn{5}{c}{} \\
    \midrule
    1     & 2     & 0.067 & \textbf{0.050} & \textbf{0.044} & 0.066 & \textbf{0.029} \\
    2     & 2     & 0.084 & 0.062 & 0.052 & 0.107 & 0.065 \\
    3     & 2     & 0.102 & 0.066 & 0.056 & 0.154 & 0.106 \\
    4     & 2     & 0.091 & 0.074 & 0.063 & 0.114 & 0.084 \\
    \midrule
    1     & 3     & 0.065 & 0.080 & 0.075 & \textbf{0.015} & \textbf{0.029} \\
    2     & 3     & 0.073 & 0.083 & 0.077 & \textbf{0.037} & 0.065 \\
    3     & 3     & 0.079 & 0.085 & 0.078 & \textbf{0.047} & 0.106 \\
    4     & 3     & 0.075 & 0.091 & 0.080 & \textbf{0.046} & 0.084 \\
    \midrule
    1     & 4     & 0.092 & 0.110 & 0.099 & \textbf{0.030} & \textbf{0.029} \\
    2     & 4     & 0.097 & 0.106 & 0.097 & 0.051 & 0.065 \\
    3     & 4     & 0.104 & 0.109 & 0.099 & 0.061 & 0.106 \\
    4     & 4     & 0.097 & 0.112 & 0.102 & 0.057 & 0.084 \\
    \midrule
    1     & 5     & 0.115 & 0.139 & 0.126 & \textbf{0.043} & \textbf{0.029} \\
    2     & 5     & 0.119 & 0.131 & 0.120 & 0.060 & 0.065 \\
    3     & 5     & 0.124 & 0.131 & 0.121 & 0.069 & 0.106 \\
    4     & 5     & 0.119 & 0.135 & 0.123 & 0.064 & 0.084 \\
    \midrule
    \end{tabular}%
    \begin{tablenotes}[para,flushleft]
		\footnotesize
		\noindent Note: This table displays the $p$-values of different backtests of the VaR and ES.  The $p$-values have been obtained based on the approach of \citet{dufour2006monte}, and those lower than $5\%$ are displayed in bold.
		\end{tablenotes}
	\end{threeparttable}
 \label{tab:pvalues_backtests_SP500}
\end{table}%

\medskip

As for the S\&P 500, the global test never rejects the null $\mathcal{H}_0$. However, in line with recommendation 2, we proceed to conduct subtests. Interestingly, each subtest yields a rejection of the null hypothesis for at least one choice of $(K, K')$, indicating potential mis-specifications in both the VaR and ES models. This underscores the importance of subtests, even when the global test fails to reject the null. Several insights emerge from this analysis. Firstly, CC VaR tests only reject the null for $K=1$ and $K'=2$, whereas the other two subtests involving Eq.\eqref{marginal1} are rejected by way more combinations of $(K, K')$. Thus, akin to the CAC40, the rejections of the null is mainly due to the severities component. Secondly, Eq. \eqref{joint4} is likely violated, given that the $p$-values of the CC VaR test are smaller than those of the CC VaR duration-based test, confirming that the null rejections primarily originate from the severities component. However, this condition is not sufficiently violated for the CC VaR test to reject the null hypothesis for other values of $K$ and $K'$. Thirdly, when condition \eqref{marginal1} is violated, condition \eqref{joint2} is automatically violated even if $H_i$ and $H_{i+1}$ are independent. Indeed, under independence, we have:
\begin{equation*}
    \mathbb{E}[Q_k(H_{i+1})Q_j(H_i)]=\mathbb{E}[Q_k(H_{i+1})]\mathbb{E}[Q_j(H_i)] \neq 0. 
\end{equation*}
This echoes the rejection of CC test of (VaR, ES) (fourth column) for a large number of combinations of $K, K'$. Fourthly, the UC test of (VaR, ES) (last column) rejects the null for all values of $K'$, but only for $K=1$. This suggests that the primary issue concerning the marginal distribution of $H_i$ is the mis-specification of its mean. 

\section{Conclusion}

In this paper, we have introduced a duration-severity approach to backtesting the ES. This new approach makes three contributions to the literature. First, it allows for the separation of the discrete component (i.e. frequency of the VaR violation) and the continuous component (severity-given-violation) of the cumulative violation process. Second, we have introduced the duration variable to the ES literature, which has the advantage of being observed synchronously with the severity variable, and hence facilitating the test of mutual independence between the duration and severity components. Third, we have derived a simple, non-parametric backtest using the theory of (bivariate) orthogonal polynomials. This approach allows us to choose the set of marginal/joint moment conditions to test freely, making it potentially more powerful than existing backtests based solely on mean and autocorrelations of the cumulative violation process. Moreover, our backtest encompasses various UC and CC tests of VaR or ES, as well as the pair (VaR, ES), as special cases. This enables the risk manager to readily identify the violated moment conditions by conducting sequential testing with nested moment conditions. 

\medskip

Like most Wald tests, our test is also two-sided. For some applications, one-sided tests could have the advantage of not penalizing overly conservative models \citep{nolde2017elicitability}. There are at least two approaches through which our test can be adapted to a one-sided test. The first one is to use general results on one-sided Wald tests \citep{silvapulle1992robust}. The second one is to rely again on orthogonal polynomials. For duration-based VaR backtests, \cite{candelon2011backtesting} assume, under the null, that the duration follows geometric distribution, but the actual coverage rate $\beta$ might be different from the required coverage level $\alpha$. They show that a simple Generalized Method of (orthogonal) Moments (GMM) allows to estimate parameter $\beta$, which can be compared with $\alpha$. Then the VaR model can be regarded as overly conservative, if $\beta> \alpha$. Similarly, for the ES component, we could assume that the severity follows a beta distribution $\mathcal{B}(\alpha_1, \alpha_2)$, which includes the uniform distribution as a special case, when $\alpha_1=\alpha_2=1$. The orthogonal polynomials associated to the beta distribution are the Jacobi polynomials \citep{gourieroux2006multivariate, ackerer2020option}. They can be used for a similar GMM estimation of the parameters $\alpha_1, \alpha_2$, and $\beta$ for the severity and duration distributions, respectively. Finally, we can compare their means, i.e., $\alpha_1/(\alpha_1+\alpha_2)$ and $\beta$ to their theoretical values $1/2$ and $\alpha$, respectively.  

\medskip

Throughout the paper, we have assumed that the validation team in charge of ES backtesting or the supervisor only uses the daily PIT compiled by the bank, but not the underlying internal model. In the case where this latter is available, it is possible to further take into account estimation risk, which leads to wider confidence intervals \citep{escanciano2010backtesting, candelon2011backtesting, lazar2019model, barendsebacktesting}. We have not followed this route since this requires the bank/trading desk to disclose their internal model, or even the trading strategy, in the case of a trading desk. As \cite{gordy2020spectral} argue, this approach preserves the objectivity and statistical integrity of the test regime by preventing the supervisor from exploiting knowledge of the internal model to bias the outcome of a test. Alternatively, when the test is to be used by the bank, it also avoids the moral hazard that the bank requests a wider confidence interval by manipulating the complexity of their internal model. Most existing studies assume that the asymptotic distribution of the estimator is normal, and in this case, \cite{candelon2011backtesting} already show how to incorporate estimation risk in a duration VaR backtest with orthogonal polynomials. However, the distribution of the model estimator could also be non-normal due to heavy-tailed phenomena \citep{su2021efficiently}, and in this case the uncertainty quantification could become much more involved. An alternative consists in transforming the moments into robust ones, i.e., moments that are orthogonal to the underlying score function as in \cite{bontemps2019moment}. This is clearly out of the scope of this paper and left for future research. 

\bibliographystyle{apalike}
\bibliography{lib}

\newpage
\appendixpage
\appendix

\section{The orthogonal polynomials}

\subsection{(Shifted) Meixner polynomials}
\label{appendix:Meixner_polynomials}

There are two definitions of the geometric distribution, depending on whether 0 belongs to the support of the distribution. Suppose we have i.i.d. Bernoulli variables with success probability $\alpha$, then the number of trials to get one success $Y \in \{1,2,...\}$ is equal to 1 plus number of failures before the first success $X \in \{0,1,2,..\}$. The geometric distribution we mentioned in Property 1 is the same as the distribution of $Y$, but the distribution of $X=Y-1$ is more popular as it is the special case of the Pascal distribution (that is a negative binomial distribution with integer number of success parameter). The orthogonal polynomials associated to this latter is called the Meixner polynomials. Then the family of orthogonal polynomial associated to the distribution of $Y$, which we call shifted Meixner polynomials, is deduced from that of $X$ by the simple change of variable $X=Y-1$. According to \cite{candelon2011backtesting}, we have:
\begin{align*}
	P_0(X, \alpha)&=1, \\
 P_1(X, \alpha)&=\frac{1-\alpha X}{\sqrt{1-\alpha}},\\
 P_{j+1}(X, \alpha)&= \frac{ (1-\alpha)(2j + 1)+\alpha (j-X + 1)}{(j+1) \sqrt{1-\alpha}} P_{j}(X, \alpha)- \frac{j}{j+1} P_{j-1}(X, \alpha), \qquad \forall j \geq 1.
\end{align*}
Thus we have: 
\begin{align*}
    P_2(X, \alpha)&= \frac{\alpha^2 x^2 +x(\alpha^2-4 \alpha )+2}{2 (1-\alpha)} \\
    P_3(X, \alpha)&=-\frac{\alpha^3 x^3 +x^2 (3 \alpha^3-9 \alpha^23)+x(2\alpha^3-9\alpha^2)+x(2 \alpha^3-9\alpha^2+18\alpha) -6}{6 (1-\alpha)^{3/2}}\\
    P_4(X, \alpha)&= \frac{1}{24} \Bigg[\frac{9 \left(\alpha^2 x (x+1)-4 \alpha x+2\right)}{\alpha-1}+\frac{(\alpha (x+3)-7) \left(\alpha^3 x \left(x^2+3 x+2\right)-9 \alpha^2 x (x+1)+18 \alpha x-6\right)}{(1-\alpha)^{2}}\Bigg].
\end{align*}

\subsection{(Shifted and normalized) Legendre polynomials}
\label{appendix:Legendre_polynomials}

The standard Legendre polynomials (see \cite{szeg1939orthogonal}), which we denote by $(\tilde{Q}_j)$, are orthogonal with respect to the uniform distribution on $(-1, 1)$. They are defined by:
\begin{align*}
    \tilde{Q}_0(X)&=1, \\
    \tilde{Q}_1(X)&= X, \\
    \tilde{Q}_{j+1}(X)&=\frac{2j+1}{j+1}X \tilde{Q}_{j}(X)- \frac{j}{j+1}\tilde{Q}_{j-1}(X), \qquad \forall j =1,2,...,
\end{align*}
and their normalized counterparts are given by $\overline{Q}_j(X)=  {\sqrt{2j+1}}\tilde{Q}_j(X), j=1...$, which satisfies:
\begin{equation}
	\label{initiallegendre}
    \frac{1}{2}\int_{-1}^1 \overline{Q}_j(x)\overline{Q}_k(x) \mathrm{d}x= \mathbbm{1}(j=k), \qquad \forall j, k.
\end{equation}

To get the orthogonal polynomials associated to the uniform distribution on $(0, 1)$, it suffices to conduct an affine change of variable $Y=2X-1$. Then from equation \eqref{initiallegendre}, we get:
\begin{equation*}
	\int_0^1 \overline{Q}_j(2y-1) \overline{Q}_k(2y-1) \mathrm{d}y=\mathbbm{1}(j=k).
\end{equation*}
Thus the orthonormal basis associated to the uniform distribution on $(0,1)$ is given by:
\begin{align*}
    Q_j(X)=\overline{Q}_j(2X-1), \qquad \forall  j.
\end{align*}
Thus we have
\begin{align*}
	Q_0(X)&=1, \\
	Q_1(X)&=\sqrt{3}(2X-1),\\
	Q_2(X)&=\sqrt{5}(6X^2-6X+1), \\
	Q_3(X)&=\sqrt{7}(20X^3-30X^2+12X{ -1}),\\
	Q_4(X)&=\frac{3}{2}(140 X^4-280 X^3{ +180 X^2-40 X+2}).
\end{align*} 

\section{Additional tables}
\label{appendix:additional_tables}

\begin{table}[H]
  \centering
  \caption{Empirical size of $5\%$ asymptotic CC VaR test based on durations}
  \resizebox{15cm}{!}{
	\renewcommand{\arraystretch}{1.6}
    \tiny
    \begin{threeparttable}
    \begin{tabular}{lcccccccc}
    \midrule
    \multicolumn{9}{c}{First DGP} \\
    \midrule
    \multicolumn{9}{c}{$\alpha$ = 0.01} \\
    \midrule
          & \multicolumn{4}{c}{$K'$ = 2}    & \multicolumn{4}{c}{$K'$ = 3} \\
    \midrule
    Sample size & \multicolumn{1}{c}{$K$ = 1} & \multicolumn{1}{c}{$K$ = 2} & \multicolumn{1}{c}{$K$ = 3} & \multicolumn{1}{c}{$K$ = 4} & \multicolumn{1}{c}{$K$ = 1} & \multicolumn{1}{c}{$K$ = 2} & \multicolumn{1}{c}{$K$ = 3} & \multicolumn{1}{c}{$K$ = 4} \\
    \midrule
    T = 250 & 0.051 & 0.057 & 0.049 & 0.048 & 0.075 & 0.061 & 0.060 & 0.053 \\
    T = 500 & 0.046 & 0.053 & 0.051 & 0.055 & 0.047 & 0.055 & 0.062 & 0.066 \\
    T = 1000 & 0.060 & 0.062 & 0.050 & 0.054 & 0.058 & 0.072 & 0.068 & 0.044 \\
    T = 2500 & 0.054 & 0.043 & 0.065 & 0.042 & 0.079 & 0.069 & 0.067 & 0.061 \\
    T = 500000 & 0.038 & 0.043 & 0.050 & 0.058 & 0.063 & 0.045 & 0.056 & 0.058 \\
    \midrule
    \multicolumn{9}{c}{$\alpha$ = 0.05} \\
    \midrule
          & \multicolumn{4}{c}{$K'$ = 2}    & \multicolumn{4}{c}{$K'$ = 3} \\
    \midrule
    Sample size & \multicolumn{1}{c}{$K$ = 1} & \multicolumn{1}{c}{$K$ = 2} & \multicolumn{1}{c}{$K$ = 3} & \multicolumn{1}{c}{$K$ = 4} & \multicolumn{1}{c}{$K$ = 1} & \multicolumn{1}{c}{$K$ = 2} & \multicolumn{1}{c}{$K$ = 3} & \multicolumn{1}{c}{$K$ = 4} \\
    \midrule
    T = 250 & 0.042 & 0.055 & 0.062 & 0.065 & 0.066 & 0.064 & 0.081 & 0.069 \\
    T = 500 & 0.057 & 0.061 & 0.045 & 0.062 & 0.057 & 0.093 & 0.045 & 0.067 \\
    T = 1000 & 0.063 & 0.067 & 0.062 & 0.048 & 0.068 & 0.068 & 0.070 & 0.053 \\
    T = 2500 & 0.042 & 0.066 & 0.052 & 0.043 & 0.065 & 0.074 & 0.076 & 0.066 \\
    T = 500000 & 0.058 & 0.056 & 0.058 & 0.043 & 0.053 & 0.055 & 0.062 & 0.056 \\
    \midrule
    \multicolumn{9}{c}{Second DGP} \\
    \midrule
    \multicolumn{9}{c}{$\alpha$ = 0.01} \\
    \midrule
          & \multicolumn{4}{c}{$K'$ = 2}    & \multicolumn{4}{c}{$K'$ = 3} \\
    \midrule
    Sample size & \multicolumn{1}{c}{$K$ = 1} & \multicolumn{1}{c}{$K$ = 2} & \multicolumn{1}{c}{$K$ = 3} & \multicolumn{1}{c}{$K$ = 4} & \multicolumn{1}{c}{$K$ = 1} & \multicolumn{1}{c}{$K$ = 2} & \multicolumn{1}{c}{$K$ = 3} & \multicolumn{1}{c}{$K$ = 4} \\
    \midrule
    T = 250 & 0.002 & 0.005 & 0.014 & 0.025 & 0.025 & 0.014 & 0.015 & 0.023 \\
    T = 500 & 0.035 & 0.021 & 0.013 & 0.012 & 0.040 & 0.030 & 0.027 & 0.028 \\
    T = 1000 & 0.048 & 0.051 & 0.054 & 0.042 & 0.048 & 0.046 & 0.053 & 0.033 \\
    T = 2500 & 0.057 & 0.055 & 0.052 & 0.057 & 0.054 & 0.068 & 0.059 & 0.069 \\
    T = 500000 & 0.041 & 0.036 & 0.048 & 0.054 & 0.060 & 0.056 & 0.063 & 0.060 \\
    \midrule
    \multicolumn{9}{c}{$\alpha$ = 0.05} \\
    \midrule
          & \multicolumn{4}{c}{$K'$ = 2}    & \multicolumn{4}{c}{$K'$ = 3} \\
    \midrule
    Sample size & \multicolumn{1}{c}{$K$ = 1} & \multicolumn{1}{c}{$K$ = 2} & \multicolumn{1}{c}{$K$ = 3} & \multicolumn{1}{c}{$K$ = 4} & \multicolumn{1}{c}{$K$ = 1} & \multicolumn{1}{c}{$K$ = 2} & \multicolumn{1}{c}{$K$ = 3} & \multicolumn{1}{c}{$K$ = 4} \\
    \midrule
    T = 250 & 0.042 & 0.061 & 0.049 & 0.042 & 0.037 & 0.053 & 0.055 & 0.063 \\
    T = 500 & 0.046 & 0.057 & 0.061 & 0.039 & 0.061 & 0.058 & 0.059 & 0.061 \\
    T = 1000 & 0.048 & 0.063 & 0.048 & 0.052 & 0.071 & 0.062 & 0.072 & 0.075 \\
    T = 2500 & 0.048 & 0.063 & 0.059 & 0.053 & 0.059 & 0.065 & 0.068 & 0.061 \\
    T = 500000 & 0.052 & 0.055 & 0.051 & 0.055 & 0.047 & 0.059 & 0.045 & 0.059 \\
    \midrule
    \end{tabular}
    	\begin{tablenotes}[para,flushleft]
		\tiny
		\noindent Note: The results are based on $1000$ replications. For each sample, we report the percentage of rejection at a $5\%$ level.
		\end{tablenotes}
	\end{threeparttable}
  \label{tab:size_alpha_005_001_CC_VaR_duration}
  }
\end{table}

\begin{table}[H]
  \centering
  \caption{Empirical size-corrected power of $5\%$ asymptotic test, $\alpha = 0.01$}
  \resizebox{15cm}{!}{
	\renewcommand{\arraystretch}{1.4}
    \tiny
    \begin{threeparttable}
    \begin{tabular}{lcccccccccc}
    \midrule
    \multicolumn{11}{c}{DGP: $A_1$} \\
    \midrule
          & \multicolumn{4}{c}{$K'$ = 2}    & \multicolumn{4}{c}{$K'$ = 3}    &       &  \\
    \midrule
    Sample size & \multicolumn{1}{c}{$K$ = 1} & \multicolumn{1}{c}{$K$ = 2} & \multicolumn{1}{c}{$K$ = 3} & \multicolumn{1}{c}{$K$ = 4} & \multicolumn{1}{c}{$K$ = 1} & \multicolumn{1}{c}{$K$ = 2} & \multicolumn{1}{c}{$K$ = 3} & \multicolumn{1}{c}{$K$ = 4} & \multicolumn{1}{l}{$BP_{ES}(5)$} & \multicolumn{1}{l}{$Z_C$} \\
    \midrule
    T = 250 & 0.015 & 0.025 & 0.021 & 0.021 & 0.022 & 0.023 & 0.030 & 0.027 &       &  \\
    T = 500 & 0.016 & 0.024 & 0.028 & 0.027 & 0.020 & 0.020 & 0.028 & 0.026 &       &  \\
    T = 1000 & 0.022 & 0.040 & 0.031 & 0.035 & 0.022 & 0.030 & 0.027 & 0.025 &       &  \\
    T = 2500 & 0.018 & 0.249 & 0.171 & 0.264 & 0.018 & 0.043 & 0.043 & 0.051 &       &  \\
    \midrule
    \multicolumn{11}{c}{DGP: $A_2$} \\
    \midrule
          & \multicolumn{4}{c}{$K'$ = 2}    & \multicolumn{4}{c}{$K'$ = 3}    &       &  \\
    \midrule
    Sample size & \multicolumn{1}{c}{$K$ = 1} & \multicolumn{1}{c}{$K$ = 2} & \multicolumn{1}{c}{$K$ = 3} & \multicolumn{1}{c}{$K$ = 4} & \multicolumn{1}{c}{$K$ = 1} & \multicolumn{1}{c}{$K$ = 2} & \multicolumn{1}{c}{$K$ = 3} & \multicolumn{1}{c}{$K$ = 4} & \multicolumn{1}{l}{$BP_{ES}(5)$} & \multicolumn{1}{l}{$Z_C$} \\
    \midrule
    T = 250 & 0.001 & 0.004 & 0.005 & 0.006 & 0.003 & 0.006 & 0.002 & 0.004 &       &  \\
    T = 500 & 0.002 & 0.000 & 0.009 & 0.009 & 0.003 & 0.002 & 0.003 & 0.003 &       &  \\
    T = 1000 & 0.003 & 0.008 & 0.011 & 0.015 & 0.002 & 0.003 & 0.002 & 0.006 &       &  \\
    T = 2500 & 0.006 & 0.010 & 0.227 & 1.000 & 0.001 & 0.003 & 0.012 & 0.079 &       &  \\
    \midrule
    \multicolumn{11}{c}{DGP: $A_3$} \\
    \midrule
          & \multicolumn{4}{c}{$K'$ = 2}    & \multicolumn{4}{c}{$K'$ = 3}    &       &  \\
    \midrule
    Sample size & \multicolumn{1}{c}{$K$ = 1} & \multicolumn{1}{c}{$K$ = 2} & \multicolumn{1}{c}{$K$ = 3} & \multicolumn{1}{c}{$K$ = 4} & \multicolumn{1}{c}{$K$ = 1} & \multicolumn{1}{c}{$K$ = 2} & \multicolumn{1}{c}{$K$ = 3} & \multicolumn{1}{c}{$K$ = 4} & \multicolumn{1}{l}{$BP_{ES}(5)$} & \multicolumn{1}{l}{$Z_C$} \\
    \midrule
    T = 250 & 0.000 & 0.000 & 0.000 & 0.000 & 0.000 & 0.000 & 0.000 & 0.000 &       &  \\
    T = 500 & 0.000 & 0.000 & 0.000 & 0.000 & 0.000 & 0.000 & 0.000 & 0.000 &       &  \\
    T = 1000 & 0.000 & 0.000 & 0.001 & 0.006 & 0.000 & 0.000 & 0.000 & 0.000 &       &  \\
    T = 2500 & 0.000 & 0.661 & 0.999 & 1.000 & 0.000 & 0.000 & 0.056 & 0.592 &       &  \\
    \midrule
    \multicolumn{11}{c}{DGP: $A_4$} \\
    \midrule
          & \multicolumn{4}{c}{$K'$ = 2}    & \multicolumn{4}{c}{$K'$ = 3}    &       &  \\
    \midrule
    Sample size & \multicolumn{1}{c}{$K$ = 1} & \multicolumn{1}{c}{$K$ = 2} & \multicolumn{1}{c}{$K$ = 3} & \multicolumn{1}{c}{$K$ = 4} & \multicolumn{1}{c}{$K$ = 1} & \multicolumn{1}{c}{$K$ = 2} & \multicolumn{1}{c}{$K$ = 3} & \multicolumn{1}{c}{$K$ = 4} & \multicolumn{1}{l}{$BP_{ES}(5)$} & \multicolumn{1}{l}{$Z_C$} \\
    \midrule
    T = 250 & 0.762 & 0.803 & 0.783 & 0.795 & 0.743 & 0.776 & 0.737 & 0.767 & 0.000 & 0.019 \\
    T = 500 & 0.939 & 0.951 & 0.938 & 0.961 & 0.894 & 0.916 & 0.935 & 0.956 & 0.038 & 0.007 \\
    T = 1000 & 0.996 & 0.999 & 0.999 & 0.999 & 0.991 & 0.990 & 0.999 & 0.996 & 0.040 & 0.004 \\
    T = 2500 & 1.000 & 1.000 & 1.000 & 1.000 & 1.000 & 1.000 & 1.000 & 1.000 & 0.030 & 0.010 \\
    \midrule
    \multicolumn{11}{c}{DGP: $A_5$} \\
    \midrule
          & \multicolumn{4}{c}{$K'$ = 2}    & \multicolumn{4}{c}{$K'$ = 3}    &       &  \\
    \midrule
    Sample size & \multicolumn{1}{c}{$K$ = 1} & \multicolumn{1}{c}{$K$ = 2} & \multicolumn{1}{c}{$K$ = 3} & \multicolumn{1}{c}{$K$ = 4} & \multicolumn{1}{c}{$K$ = 1} & \multicolumn{1}{c}{$K$ = 2} & \multicolumn{1}{c}{$K$ = 3} & \multicolumn{1}{c}{$K$ = 4} & \multicolumn{1}{l}{$BP_{ES}(5)$} & \multicolumn{1}{l}{$Z_C$} \\
    \midrule
    T = 250 & 0.041 & 0.059 & 0.048 & 0.040 & 0.077 & 0.060 & 0.036 & 0.068 & 0.000 & 0.047 \\
    T = 500 & 0.081 & 0.051 & 0.062 & 0.084 & 0.057 & 0.086 & 0.056 & 0.073 & 0.112 & 0.050 \\
    T = 1000 & 0.090 & 0.087 & 0.114 & 0.126 & 0.101 & 0.114 & 0.108 & 0.138 & 0.151 & 0.081 \\
    T = 2500 & 0.140 & 0.144 & 0.166 & 0.145 & 0.165 & 0.150 & 0.139 & 0.145 & 0.211 & 0.071 \\
    \midrule
    \end{tabular}%
      	\begin{tablenotes}[para,flushleft]
		\tiny
		\noindent Note: The results are based on $1000$ replications and are size-corrected. For each sample, we report the percentage of rejection at a $5\%$ level. The conditional test statistic of \citet{du2017backtesting} is denoted as $BP_{ES}(m)$, where $m$ is the degree of freedom, and while $Z_C$ corresponds to the conditional test statistic of \citet{acerbi2014back}.
		\end{tablenotes}
	\end{threeparttable}
  \label{tab:power_alpha_001}
  }
\end{table}

\begin{table}[H]
  \centering
  \caption{Empirical size-corrected power of $5\%$ asymptotic CC VaR test based on durations, $\alpha = 0.05$}
  \resizebox{15cm}{!}{
	\renewcommand{\arraystretch}{1.4}
    \tiny
    \begin{threeparttable}
    \begin{tabular}{lcccccccc}
    \midrule
    \multicolumn{9}{c}{DGP: $A_1$} \\
    \midrule
          & \multicolumn{4}{c}{$K'$ = 2}    & \multicolumn{4}{c}{$K'$ = 3} \\
    \midrule
    \multicolumn{1}{l}{Sample size} & \multicolumn{1}{c}{$K$ = 1} & \multicolumn{1}{c}{$K$ = 2} & \multicolumn{1}{c}{$K$ = 3} & \multicolumn{1}{c}{$K$ = 4} & \multicolumn{1}{c}{$K$ = 1} & \multicolumn{1}{c}{$K$ = 2} & \multicolumn{1}{c}{$K$ = 3} & \multicolumn{1}{c}{$K$ = 4} \\
    \midrule
    T = 250 & 0.050 & 0.050 & 0.050 & 0.050 & 0.050 & 0.050 & 0.050 & 0.050 \\
    T = 500 & 0.050 & 0.050 & 0.050 & 0.050 & 0.050 & 0.050 & 0.050 & 0.050 \\
    T = 1000 & 0.050 & 0.050 & 0.050 & 0.050 & 0.050 & 0.050 & 0.050 & 0.050 \\
    T = 2500 & 0.050 & 0.050 & 0.050 & 0.050 & 0.050 & 0.050 & 0.050 & 0.050 \\
    \midrule
    \multicolumn{9}{c}{DGP: $A_2$} \\
    \midrule
          & \multicolumn{4}{c}{$K'$ = 2}    & \multicolumn{4}{c}{$K'$ = 3} \\
    \midrule
    \multicolumn{1}{l}{Sample size} & \multicolumn{1}{c}{$K$ = 1} & \multicolumn{1}{c}{$K$ = 2} & \multicolumn{1}{c}{$K$ = 3} & \multicolumn{1}{c}{$K$ = 4} & \multicolumn{1}{c}{$K$ = 1} & \multicolumn{1}{c}{$K$ = 2} & \multicolumn{1}{c}{$K$ = 3} & \multicolumn{1}{c}{$K$ = 4} \\
    \midrule
    T = 250 & 0.000 & 0.001 & 0.033 & 0.074 & 0.000 & 0.000 & 0.000 & 0.000 \\
    T = 500 & 0.000 & 0.015 & 0.977 & 0.999 & 0.000 & 0.000 & 0.655 & 0.877 \\
    T = 1000 & 0.000 & 0.819 & 1.000 & 1.000 & 0.000 & 0.129 & 1.000 & 1.000 \\
    T = 2500 & 0.000 & 1.000 & 1.000 & 1.000 & 0.000 & 1.000 & 1.000 & 1.000 \\
    \midrule
    \multicolumn{9}{c}{DGP: $A_3$} \\
    \midrule
          & \multicolumn{4}{c}{$K'$ = 2}    & \multicolumn{4}{c}{$K'$ = 3} \\
    \midrule
    \multicolumn{1}{l}{Sample size} & \multicolumn{1}{c}{$K$ = 1} & \multicolumn{1}{c}{$K$ = 2} & \multicolumn{1}{c}{$K$ = 3} & \multicolumn{1}{c}{$K$ = 4} & \multicolumn{1}{c}{$K$ = 1} & \multicolumn{1}{c}{$K$ = 2} & \multicolumn{1}{c}{$K$ = 3} & \multicolumn{1}{c}{$K$ = 4} \\
    \midrule
    T = 250 & 0.000 & 0.000 & 0.024 & 0.063 & 0.000 & 0.000 & 0.000 & 0.000 \\
    T = 500 & 0.000 & 0.016 & 0.976 & 1.000 & 0.000 & 0.000 & 0.668 & 0.870 \\
    T = 1000 & 0.000 & 0.813 & 1.000 & 1.000 & 0.000 & 0.098 & 1.000 & 1.000 \\
    T = 2500 & 0.000 & 1.000 & 1.000 & 1.000 & 0.000 & 1.000 & 1.000 & 1.000 \\
    \midrule
    \multicolumn{9}{c}{DGP: $A_4$} \\
    \midrule
          & \multicolumn{4}{c}{$K'$ = 2}    & \multicolumn{4}{c}{$K'$ = 3} \\
    \midrule
    \multicolumn{1}{l}{Sample size} & \multicolumn{1}{c}{$K$ = 1} & \multicolumn{1}{c}{$K$ = 2} & \multicolumn{1}{c}{$K$ = 3} & \multicolumn{1}{c}{$K$ = 4} & \multicolumn{1}{c}{$K$ = 1} & \multicolumn{1}{c}{$K$ = 2} & \multicolumn{1}{c}{$K$ = 3} & \multicolumn{1}{c}{$K$ = 4} \\
    \midrule
    T = 250 & 0.360 & 0.228 & 0.200 & 0.235 & 0.225 & 0.145 & 0.118 & 0.071 \\
    T = 500 & 0.641 & 0.482 & 0.417 & 0.443 & 0.344 & 0.314 & 0.236 & 0.245 \\
    T = 1000 & 0.937 & 0.859 & 0.830 & 0.770 & 0.689 & 0.684 & 0.539 & 0.484 \\
    T = 2500 & 1.000 & 1.000 & 1.000 & 1.000 & 0.997 & 0.998 & 0.991 & 0.986 \\
    \midrule
    \multicolumn{9}{c}{DGP: $A_5$} \\
    \midrule
          & \multicolumn{4}{c}{$K'$ = 2}    & \multicolumn{4}{c}{$K'$ = 3} \\
    \midrule
    \multicolumn{1}{l}{Sample size} & \multicolumn{1}{c}{$K$ = 1} & \multicolumn{1}{c}{$K$ = 2} & \multicolumn{1}{c}{$K$ = 3} & \multicolumn{1}{c}{$K$ = 4} & \multicolumn{1}{c}{$K$ = 1} & \multicolumn{1}{c}{$K$ = 2} & \multicolumn{1}{c}{$K$ = 3} & \multicolumn{1}{c}{$K$ = 4} \\
    \midrule
    T = 250 & 0.128 & 0.128 & 0.147 & 0.142 & 0.141 & 0.114 & 0.141 & 0.101 \\
    T = 500 & 0.156 & 0.176 & 0.198 & 0.224 & 0.161 & 0.176 & 0.179 & 0.173 \\
    T = 1000 & 0.209 & 0.256 & 0.306 & 0.273 & 0.213 & 0.291 & 0.231 & 0.228 \\
    T = 2500 & 0.317 & 0.434 & 0.473 & 0.418 & 0.326 & 0.413 & 0.391 & 0.415 \\
    \midrule
    \end{tabular}%
      	\begin{tablenotes}[para,flushleft]
		\tiny
		\noindent Note: The results are based on $1000$ replications and are size-corrected. For each sample, we report the percentage of rejection at a $5\%$ level.
		\end{tablenotes}
	\end{threeparttable}
  \label{tab:power_alpha_005_CC_VaR_duration}
  }
\end{table}

\begin{table}[H]
  \centering
  \caption{Empirical size-corrected power of $5\%$ asymptotic CC VaR test based on durations, $\alpha = 0.01$}
  \resizebox{15cm}{!}{
	\renewcommand{\arraystretch}{1.4}
    \tiny
    \begin{threeparttable}
    \begin{tabular}{lcccccccc}
    \midrule
    \multicolumn{9}{c}{DGP: $A_1$} \\
    \midrule
          & \multicolumn{4}{c}{$K'$ = 2}    & \multicolumn{4}{c}{$K'$ = 3} \\
    \midrule
    Sample size & \multicolumn{1}{c}{$K$ = 1} & \multicolumn{1}{c}{$K$ = 2} & \multicolumn{1}{c}{$K$ = 3} & \multicolumn{1}{c}{$K$ = 4} & \multicolumn{1}{c}{$K$ = 1} & \multicolumn{1}{c}{$K$ = 2} & \multicolumn{1}{c}{$K$ = 3} & \multicolumn{1}{c}{$K$ = 4} \\
    \midrule
    T = 250 & 0.050 & 0.050 & 0.050 & 0.050 & 0.050 & 0.050 & 0.050 & 0.050 \\
    T = 500 & 0.050 & 0.050 & 0.050 & 0.050 & 0.050 & 0.050 & 0.050 & 0.050 \\
    T = 1000 & 0.050 & 0.050 & 0.050 & 0.050 & 0.050 & 0.050 & 0.050 & 0.050 \\
    T = 2500 & 0.050 & 0.050 & 0.050 & 0.050 & 0.050 & 0.050 & 0.050 & 0.050 \\
    \midrule
    \multicolumn{9}{c}{DGP: $A_2$} \\
    \midrule
          & \multicolumn{4}{c}{$K'$ = 2}    & \multicolumn{4}{c}{$K'$ = 3} \\
    \midrule
    Sample size & \multicolumn{1}{c}{$K$ = 1} & \multicolumn{1}{c}{$K$ = 2} & \multicolumn{1}{c}{$K$ = 3} & \multicolumn{1}{c}{$K$ = 4} & \multicolumn{1}{c}{$K$ = 1} & \multicolumn{1}{c}{$K$ = 2} & \multicolumn{1}{c}{$K$ = 3} & \multicolumn{1}{c}{$K$ = 4} \\
    \midrule
    T = 250 & 0.000 & 0.000 & 0.000 & 0.000 & 0.000 & 0.000 & 0.000 & 0.000 \\
    T = 500 & 0.000 & 0.000 & 0.000 & 0.000 & 0.000 & 0.000 & 0.000 & 0.000 \\
    T = 1000 & 0.000 & 0.000 & 0.000 & 0.006 & 0.000 & 0.000 & 0.000 & 0.000 \\
    T = 2500 & 0.000 & 0.036 & 1.000 & 1.000 & 0.000 & 0.000 & 0.987 & 1.000 \\
    \midrule
    \multicolumn{9}{c}{DGP: $A_3$} \\
    \midrule
          & \multicolumn{4}{c}{$K'$ = 2}    & \multicolumn{4}{c}{$K'$ = 3} \\
    \midrule
    Sample size & \multicolumn{1}{c}{$K$ = 1} & \multicolumn{1}{c}{$K$ = 2} & \multicolumn{1}{c}{$K$ = 3} & \multicolumn{1}{c}{$K$ = 4} & \multicolumn{1}{c}{$K$ = 1} & \multicolumn{1}{c}{$K$ = 2} & \multicolumn{1}{c}{$K$ = 3} & \multicolumn{1}{c}{$K$ = 4} \\
    \midrule
    T = 250 & 0.000 & 0.000 & 0.000 & 0.000 & 0.000 & 0.000 & 0.000 & 0.000 \\
    T = 500 & 0.000 & 0.000 & 0.000 & 0.000 & 0.000 & 0.000 & 0.000 & 0.000 \\
    T = 1000 & 0.000 & 0.000 & 0.000 & 0.006 & 0.000 & 0.000 & 0.000 & 0.000 \\
    T = 2500 & 0.000 & 0.035 & 1.000 & 1.000 & 0.000 & 0.000 & 0.991 & 1.000 \\
    \midrule
    \multicolumn{9}{c}{DGP: $A_4$} \\
    \midrule
          & \multicolumn{4}{c}{$K'$ = 2}    & \multicolumn{4}{c}{$K'$ = 3} \\
    \midrule
    Sample size & \multicolumn{1}{c}{$K$ = 1} & \multicolumn{1}{c}{$K$ = 2} & \multicolumn{1}{c}{$K$ = 3} & \multicolumn{1}{c}{$K$ = 4} & \multicolumn{1}{c}{$K$ = 1} & \multicolumn{1}{c}{$K$ = 2} & \multicolumn{1}{c}{$K$ = 3} & \multicolumn{1}{c}{$K$ = 4} \\
    \midrule
    T = 250 & 0.859 & 0.787 & 0.773 & 0.738 & 0.724 & 0.777 & 0.746 & 0.680 \\
    T = 500 & 0.934 & 0.937 & 0.929 & 0.940 & 0.893 & 0.928 & 0.916 & 0.908 \\
    T = 1000 & 0.999 & 0.995 & 0.991 & 0.992 & 0.994 & 0.988 & 0.989 & 0.992 \\
    T = 2500 & 1.000 & 1.000 & 1.000 & 1.000 & 1.000 & 1.000 & 1.000 & 1.000 \\
    \midrule
    \multicolumn{9}{c}{DGP: $A_5$} \\
    \midrule
          & \multicolumn{4}{c}{$K'$ = 2}    & \multicolumn{4}{c}{$K'$ = 3} \\
    \midrule
    Sample size & \multicolumn{1}{c}{$K$ = 1} & \multicolumn{1}{c}{$K$ = 2} & \multicolumn{1}{c}{$K$ = 3} & \multicolumn{1}{c}{$K$ = 4} & \multicolumn{1}{c}{$K$ = 1} & \multicolumn{1}{c}{$K$ = 2} & \multicolumn{1}{c}{$K$ = 3} & \multicolumn{1}{c}{$K$ = 4} \\
    \midrule
    T = 250 & 0.052 & 0.063 & 0.070 & 0.076 & 0.066 & 0.067 & 0.084 & 0.083 \\
    T = 500 & 0.072 & 0.090 & 0.081 & 0.109 & 0.075 & 0.106 & 0.081 & 0.074 \\
    T = 1000 & 0.102 & 0.096 & 0.108 & 0.108 & 0.122 & 0.132 & 0.114 & 0.139 \\
    T = 2500 & 0.123 & 0.149 & 0.164 & 0.163 & 0.159 & 0.152 & 0.171 & 0.151 \\
    \midrule
    \end{tabular}%
      	\begin{tablenotes}[para,flushleft]
		\tiny
		\noindent Note: The results are based on $1000$ replications and are size-corrected. For each sample, we report the percentage of rejection at a $5\%$ level.
		\end{tablenotes}
	\end{threeparttable}
  \label{tab:power_alpha_001_CC_VaR_duration}
  }
\end{table}

\begin{table}[H]
  \centering
  \caption{Estimated parameters}
  \begin{threeparttable}
    \begin{tabular}{lcc}
    \midrule
          & \multicolumn{1}{l}{CAC40} & \multicolumn{1}{l}{S\&P500} \\
    \midrule
    $\delta_0$ & 0.073 & 0.085 \\
    $\delta_1$ & 0.032 & -0.030 \\
    $\gamma_0$ & 0.058 & 0.017 \\
    $\gamma_1$ & 0.162 & 0.174 \\
    $\gamma_2$ & 0.766 & 0.825 \\
    $\nu$    & 5.000 & 4.000 \\
    $F^{-1}_\nu(\alpha)$ & -2.015 & -2.132 \\
    $\mu(\alpha)$ & -2.800 & -2.601 \\
    \midrule
    \end{tabular}
    \begin{tablenotes}[para,flushleft]
		\footnotesize
		\noindent Note: This table displays the values of parameters estimated on the in-sample period.
		\end{tablenotes}
	\end{threeparttable}
 \label{tab:estimated_model}
\end{table}

\begin{table}[!htbp]
  \centering
  \caption{Ljung-Box Test}
  \begin{threeparttable}
    \begin{tabular}{lcc}
    \midrule
          & \multicolumn{1}{l}{CAC40} & \multicolumn{1}{l}{S\&P500} \\
    \midrule
    Stand. Innov. & 0.694 & 0.388 \\
    Sq. Stand. Innov. & 0.929 & 0.916 \\
    \midrule
    \end{tabular}
      \begin{tablenotes}[para,flushleft]
		\footnotesize
		\noindent Note: This table displays the $p$-values of Ljung-Box tests applied to the in-sample, raw and squared standardized residuals. The $p$-values suggest that the model appropriately captures the return dynamics.
		\end{tablenotes}
	\end{threeparttable}
 \label{tab:LB_test}
\end{table}

\end{document}